\documentclass[oneside,pdflatex,sn-mathphys-num]{sn-jnl}

\usepackage{multirow}%
\usepackage{amsmath,amssymb,amsfonts}%
\usepackage{amsthm}%
\usepackage{mathrsfs}%
\usepackage[title]{appendix}%
\usepackage{xcolor}%
\usepackage{manyfoot}%
\usepackage{booktabs}%
\usepackage{algorithm}%
\usepackage{algorithmicx}%
\usepackage{algpseudocode}%
\usepackage{listings}%
\usepackage{hyperref}
\usepackage{natbib}
\usepackage{subcaption}
\usepackage{caption}
\usepackage[dvipsnames]{xcolor}
\usepackage[normalem]{ulem}
\usepackage{ulem}
\usepackage{soul}
\usepackage{cancel}

\usepackage{makecell}
\usepackage{booktabs}
\usepackage{multirow}
\usepackage{graphicx}   
\usepackage{textcomp}   
\usepackage{pdflscape}
\usepackage{geometry}
\usepackage{etoolbox}
\usepackage{tikz}
\usepackage{enumitem}
\usepackage{float} 
\usepackage{subcaption}
\usetikzlibrary{arrows.meta, positioning}
\usepackage{tikz}
\usetikzlibrary{positioning, decorations.pathreplacing}

\makeatletter
\@twosidefalse
\@mparswitchfalse
\makeatother

\theoremstyle{thmstyleone}%
\newtheorem{theorem}{Theorem}
\theoremstyle{thmstyletwo}%
\newtheorem{remark}{Remark}%

\theoremstyle{thmstylethree}%
\newtheorem{definition}{Definition}%

\begin{document}

\title[Article Title]{A New Broadcast Model for Several Network Topologies}


\author[1]{\fnm{Hongbo} \sur{Lu}}

\author[1,2]{\fnm{Junsung} \sur{Hwang}}
\equalcont{These authors contributed equally to this work.}

\author[1]{\fnm{Bernard} \sur{Tenreiro}}
\equalcont{These authors contributed equally to this work.}

\author[1]{\fnm{Nabila Jaman} \sur{Tripti}}

\author[1,2]{\fnm{Darren} \sur{Hamilton}}

\author*[1]{\fnm{Yuefan} \sur{Deng}}\email{yuefan.deng@stonybrook.edu}

\affil[1]{\orgdiv{Department of Applied Mathematics and Statistics}, \orgname{Stony Brook University}, \orgaddress{\city{Stony Brook}, \postcode{11794}, \state{NY}, \country{USA}}}
\affil[2]{\orgdiv{Department of Computer Science}, \orgname{Stony Brook University}, \orgaddress{\city{Stony Brook}, \postcode{11794}, \state{NY}, \country{USA}}}


\abstract{We introduce Broadcast by Balanced Saturation (BBS), a general class of tree-based pipelined broadcast algorithms that optimizes communication efficiency across diverse network topologies, with a particular emphasis on large message sizes. By addressing spanning tree construction and communication task scheduling, two fundamental theoretical challenges in broadcasting, BBS offers a unified and flexible framework that operates effectively under varied network constraints. The algorithm maximizes aggregated throughput while simultaneously addressing topology constraints, synchronization overhead, bandwidth limitations and contention. Using SimGrid under standard assumptions, including full-duplex and one-port communication, various algorithms were evaluated on Mesh, Butterfly, Dragonfly, and Fat-Tree topologies. Results demonstrate that BBS consistently outperforms both general-purpose and topology-aware broadcast algorithms across a wide range of topologies and message sizes, establishing it as a robust and high-performance solution for large-scale systems.}




\maketitle
\section{Introduction}\label{sec1}
In the age of supercomputing, optimal performance is a necessity for the advancement of scientific fields, particularly in big data domains such as artificial intelligence, molecular dynamics, and climate modeling, as well as industrial applications including media, energy, financial, and information technology \cite{almeida2025assessing, 9355242, doi:10.1177/10943420231183688, doi:10.1021/ct9000685, app10196717}. One fundamental task in distributed systems and parallel computing is broadcasting, in which a source node contains a message that must be received by all other nodes in a network. Broadcasting is necessary for a wide range of parallelized tasks, including matrix operations, shortest paths and other graph operations, and other algorithms, namely the Fast Fourier Transform \cite{10.5555/899255, 10.1007/11549468_87, dongarra2003linpack}.

Many broadcast algorithms have been proposed that are deterministic \cite{HASANOV201530, article1212,fat_tree_alg, 10.1007/978-3-540-30218-6_28, 10.1007/11557654_8, 1420226, novel_pipelined, pipeline_bcast, twotree, Bine, article_pfDeng, GLF, GLF2} and stochastic \cite{8361902, 10.1145/1400751.1400773}, which are applied to different domains depending on the nature of the network topology. In supercomputers and high-performance computing clusters, where the network topology is known, a deterministic broadcast algorithm guarantees performance predictability. Broadcast implementations using the Message Passing Interface (MPI), a standardized protocol for parallel computing, employ general broadcast algorithms including binomial tree, flat tree, pipeline, and scatter-allgather. The actual algorithm is dynamically selected based on factors such as message size and the number of processors \cite{MPICH, f7a8565e2a44463d956da9c22c088375}.

Many earlier supercomputers utilize simple network topologies such as mesh, hypercube, and torus \cite{679219, blue_gene, ring_torus_hyper, compare_old_super}. However, modern supercomputers utilize specialized topologies, including Butterfly \cite{flat_butterfly}, Dragonfly \cite{dragonfly}, and Fat-Tree \cite{9926123, fat_tree}, which may include a high radix and connection density, low diameter, and other special properties \cite{6337589} for transmission optimization. Such topologies are often hierarchical topologies, such as where nodes connect to routers, which then form a network. In such settings, relying on general broadcast algorithms may lead to suboptimal execution times, so specialized algorithms are often considered \cite{MPI_bluegene, top_aware_infini, fat_tree_alg, article_pfDeng, GLF, GLF2}. A general-purpose, efficient deterministic broadcast algorithm that can adapt to any topology and communication assumptions is both incredibly practical and of theoretical interest. One common class of broadcast is the pipelined broadcast, where the message is partitioned and sent independently, which can be more efficient in certain conditions. This class of broadcasting will be the focus of this work. 

A network topology is initially treated as an undirected graph, where each node represents a computing unit and an edge between nodes represents a communication connection between two computing units. This work often models the underlying topology as a directed graph to better distinguish sending and receiving operations and to define communication protocols. In general, each topology will also have communication rules, such as full or half-duplex. This work will address different communication rules when necessary. This setting is consistent with previous work on deterministic distributed scheduling, and captures common assumptions in message-passing environments, including MPI-based clusters, where collective operations are coordinated across known topologies.

This work introduces a general class of broadcast algorithms without relying on any strong structural assumptions. Our key contributions are as follows:
\begin{enumerate}[label=\arabic*)]
    \item Universality: This work proposes a general class of broadcast algorithms and theoretical foundations applicable to diverse network topologies under various communication constraints.
    
    \item Performance: Bounds on building the algorithm and bounds on executing the algorithm are provided to ensure feasibility and performance guarantees.

    \item Low storage: A specific algorithm can be built for a topology once offline or at the initialization time, and the schedule is lightweight.

    \item Evaluation: Under common communication constraints, we use SimGrid simulations to evaluate broadcast algorithms across different network topologies and message sizes.
\end{enumerate}

Together, these contributions form a foundation for predictable and topology-aware broadcasting in distributed systems, with direct applications to scheduling on high-performance computing platforms.

\section{Related Work}\label{related_work}
Broadcast communication has been extensively studied in parallel and distributed systems, with a wide range of algorithms proposed to balance latency, bandwidth utilization, and topology conditions. Classical approaches are largely based on structured communication patterns such as binomial trees, which remain a common baseline in MPI implementations such as MPICH \cite{MPICH}. In these structures, the number of participating processes doubles at each step, yielding logarithmic latency, well-suited for small messages where latency is the dominant factor. However, binomial trees underutilize available bandwidth for large messages, as communication is concentrated among interior nodes, leaving leaf nodes idle. 

To address these limitations, pipelined broadcast algorithms partition messages into smaller segments to enable concurrent transmission. A prominent example is the Van de Geijn algorithm, which decomposes the broadcast into a scatter phase followed by an allgather phase, achieving high throughput for large messages \cite{MPICH}. Further proposed advancements include multi-tree factorizations, such as the 2Tree algorithm, which exploits full-duplex communication by distributing data simultaneously over two binary trees \cite{twotree}. By scheduling communication so that leaf nodes in one tree correspond to interior nodes in the other, the 2Tree algorithm can double the achieved bandwidth compared to a binomial tree. Another approach, the LockStep Broadcast Tree algorithm, optimizes pipelining for heterogeneous networks by defining a basic unit of upload bandwidth to disseminate data chunks across nodes with varying capacities \cite{novel_pipelined}. Despite the availability of these diverse algorithms, MPI implementations typically rely on empirically tuned decision rules that may fail to select the optimal algorithm on certain platforms, leading to significant performance degradation. One approach employs a performance model to select the most suitable broadcast algorithm based on factors such as communication protocol, process count, and message size \cite{nuriyev2022model}. Accurate selection, however, requires detailed analytical models that differentiate between shared-memory and network communication costs and account for the effects of network congestion.

Consequently, recent research has shifted toward topology-aware algorithms that exploit specialized network structures to enhance performance. For example, Bine (binomial negabinary) trees are utilized for broadcasting on hierarchical topologies as a general advancement over binomial trees \cite{Bine}. Unlike traditional trees that minimize physical rank distance, Bine trees minimize the modular distance between communicating processes. This approach naturally optimizes for hierarchical networks, such as Dragonfly and Fat-Trees, by ensuring that communication remains within local, fully connected groups as much as possible, reducing traffic on oversubscribed global links. Other approaches tailor broadcast algorithms to hierarchical topologies, including specialized routing strategies. For Dragonfly networks, algorithms employ specialized routing schedules to manage the local and global link traversal \cite{GLF, GLF2}. On Fat-Tree networks, research has explored multi-color k-ary trees that map parent nodes across different trees to disjoint subsets of physical nodes to maximize link utilization \cite{fat_tree_alg}. While these methods achieve high performance on their target systems, they require specific network conditions and do not provide general frameworks.

Complementing these empirical designs, theoretical research has modeled broadcasting as a throughput maximization problem on weighted directed graphs in non-uniform environments \cite{pipeline_bcast}. Their approach decomposes communication into a set of concurrently used spanning trees derived via linear programming, and is asymptotically optimal for an infinite sequence of messages. Despite its theoretical optimality, the approach has practical limitations. This formulation targets steady-state throughput rather than the completion time of a single broadcast, and assumes a single message can be split into an infinite number of packets. Consequently, it may not be effective for minimizing the completion time of a single, finite-sized broadcast.

\section{The Broadcast Problem}\label{bcast_model}
\subsection{Background}
This work defines the broadcast problem as follows. In a given directed topology $\vec{G}=(V,\vec{E})$ and a given single root node containing a message of size $m$, the problem is to construct an algorithm such that all other $|V|-1$ nodes receive the message as fast as possible. Data is assumed to be transmitted via physical or virtual links between nodes as directed edges that follow certain intersection rules depending on network assumptions, such as full or half-duplex. Data cannot flow through intersecting edges simultaneously. Under network assumptions, an intersection graph $G_I$ can be constructed from the underlying topology $\vec{G}$, which specifies which directed edge can be simultaneously utilized with all other directed edges. Using this intersection graph, special communication protocols such as link hopping are allowed, but there must be a defined directed edge in $\vec{E}$ with specified intersecting information stored in $G_I$.

Before designing and optimizing a broadcast algorithm, the communication costs must be defined. This work uses a simple model to estimate the time of the communication from node $i$ to node $j$ via their connection $(i,j)$. This connection, which is referred to as a directed edge, has latency $L_{i,j}$ and bandwidth $B_{i,j}$. The total time to send data of size $n$ along the directed edge $(i,j)$ is given by $L_{i,j}+\frac{n}{B_{i,j}}$. This time model is very similar to the classic Hockney model \cite{hock_model}, without assuming uniform bandwidths and latencies. The Hockney model is the basis for many cost analyses of collective operations; however, more general models exist, such as LogP and its extensions \cite{logp, loggp, plogp, loggps}, which aim to more realistically model parallel computation.

\subsection{Broadcast Algorithms}\label{bcast_problem}
A broadcast algorithm must provide time-dependent instructions specifying when two nodes communicate data, the amount of data transmitted, and the direction of transmission. Among all broadcast algorithms, there is a special class commonly referred to as pipelined broadcast algorithms.

\begin{definition} 
A broadcast algorithm is \emph{pipelined} if the message is partitioned into fixed packets, and this partition does not vary throughout the algorithm execution. Each node is only responsible for forwarding entire packets to its neighbors, without modifying packet contents or sizes.
\end{definition}

Similar to other works, this work focuses on the analysis of this definition of pipelined broadcast algorithms for implementation, performance, and analysis purposes. Permitting individual nodes to repartition the message would not only introduce unnecessary overhead, but also introduce a great amount of implementation challenges that would otherwise not exist for this definition of pipelined broadcast algorithms. Another class of broadcast algorithms is referred to as tree-based, where data is sent along spanning trees of the underlying topology.

\begin{definition} 
A pipelined broadcast algorithm is \emph{tree-based} if the communication pattern for each packet follows a directed spanning tree of the underlying graph. Different packets may be routed along different spanning trees.
\end{definition}

\begin{theorem}
Every pipelined broadcast algorithm can be formulated as a tree-based pipelined broadcast algorithm.
\end{theorem}

\begin{proof}
Consider an arbitrary packet in a pipelined broadcast algorithm. Since nodes forward unmodified packets, the propagation of this packet can be uniquely traced through the sequence of nodes that first receive it. Construct a directed graph whose directed edges represent transmissions of this packet, and consider the first occurrence at which each node receives the packet. This subgraph must contain no cycles, since the packet is received only once per node, and is therefore a tree rooted at the source node. To complete the broadcast, every node must eventually receive the packet. Hence, this tree spans all nodes, thus forming a spanning tree of the underlying graph. 

Since this argument holds for every packet, each packet is sent along a spanning tree, which may differ per packet. Therefore, any pipelined broadcast algorithm can be formulated as tree-based.
\end{proof}

\begin{remark}
A tree-based pipelined algorithm formulated from a pipelined algorithm must perform at least as well. This is because a pipelined broadcast algorithm may contain cycles, where the same data is repeatedly received by some node. In this case, the tree-based algorithm will remove all redundant communications.
\end{remark}

By the earlier definition, a broadcast algorithm must provide time-dependent instructions specifying communication. This definition can be formulated as a function of time and directional edges, regardless of communication constraints and assumptions.

\begin{definition} 
A pipelined broadcast algorithm $\mathcal{A}$ can be formally defined as a function of time and the directional edges $\vec{E}$ in the underlying graph $\vec{G}$ as:
\begin{equation}
\mathcal{A}: [0, \infty) \times \vec{E} \to \{0, \ 1\}
\label{bcast_alg_def}
\end{equation}
where the operator $\times$ denotes the Cartesian product.
\end{definition} 

At each time instance in the interval $[0, \infty)$, $\mathcal{A}$ maps each directed edge to $\{0, 1\}$, where 0 indicates the directed edge as not active and 1 as active. An active directed edge $(i,j)$ corresponds to the node $i$ sending data to the node $j$. The mapping must avoid conflicting edges, as specified by the intersection graph $G_I$. Such conflicts may occur due to communication rules, such as full or half-duplex and physical link sharing. As packets cannot be repartitioned once they are in transit, they must be fully sent without interruption. Equation \ref{bcast_alg_def} fully defines a pipelined broadcast algorithm. It specifies when two nodes communicate data and the direction of transmission by labeling which directed edges are active. The amount of data transmitted is defined by the time interval when the directed edge is active, assuming a known rate of transmission per edge. A tree-based pipelined broadcast algorithm can similarly be defined. 
\begin{definition}
Given $K$ trees $\{T_k\}_{k=1}^K$, and the collection of their directed edges $\{\vec{E}(T_k)\}_{k=1}^K$, a tree-based pipelined broadcast algorithm $\mathcal{A}_{p}$ is defined as:
\[
\mathcal{A}_p: [0, \infty) \times \{\vec{E}(T_k)\}_{k=1}^K \to \{0, \ 1\}.
\label{tree_pipe_alg_def}
\]
which fully defines a pipelined tree-based broadcast algorithm, prohibits message repartitioning, and similarly avoids conflicts defined in $G_I$.
\end{definition}

To actually build a pipelined tree-based broadcast algorithm, the same structure must always be constructed. First, for each packet, there must exist a spanning tree from the root. Second, each node must have a local ordered schedule that instructs the node what to do, specifically as an ordered list of send/receive pairs. The first task of constructing a set of spanning trees from the root is known as the spanning arborescence problem, and the second task is known as the task scheduling problem. Adopting these notations, to build a pipelined tree-based broadcast algorithm, both an arborescence solver and a schedule solver are necessary. It is important to note that simple existence results for both problems are trivial for a physically connected topology. However, such results will produce a finishing time that is far worse than that of at least one other known suboptimal algorithm. Thus, suitable objectives, derived from the finishing time, are necessary for each solver.

\subsection{Performance Analysis}\label{time_profile}
For any pipelined tree-based broadcast algorithm with a collection of $K$ spanning trees, its performance can be evaluated based on its time of completion. For an algorithm, the finishing time as a function of the message size is called the time profile. The time profile can be defined for any pipelined broadcast algorithm, with any general assumptions, for performance comparison between algorithms.

\begin{definition}
The total message size $M$ is partitioned into $m$ groups of packets. Each group of packets consists of $\{P_k\}_{k=1}^K$ individual packets, with each packet $P_k$ being of size $M_k$, which will be sent along one of the $K$ trees. Only the last group is allowed to have packets strictly smaller than $M_k$ or to have empty packets. Lastly, the particular schedule for any two groups of packets only differs by a translation in time.
\end{definition}

Under this definition, two trees $T_i$ and $T_j$ are allowed to be the same tree structurally but carry different packets $P_i$ and $P_j$. Each tree $T_k$ defines the communication pattern of one packet with size $P_k$, and all trees follow a pipelined schedule. Such a schedule cycles through different groups to finish the broadcast of the message. 
\begin{definition}
Under this structure, denote the time it takes to complete a broadcast with message size $M$ as $T(M)$. This function is called the time profile of a broadcast algorithm.
\end{definition}
\begin{theorem}
For any pipelined broadcast algorithm with $M$ evenly divided by $\sum_k M_k$, the time profile $T(M)$ satisfies
\begin{equation}
T(M)=a+b\frac{M}{\sum_k M_k},
\label{time_prof_eq}
\end{equation}
where $a$ and $b$ depend on the $K$ spanning trees and the broadcast schedule.
\end{theorem}

\begin{proof}
The quantity $\sum_k M_k$ denotes the size of each group of the $m$ groups of packets, and $T(\sum_k M_k)$ is the time the algorithm takes to complete the broadcast of a single group of packets. In the broadcast of any single group of packets with 0 starting time, for each node $i$, take the start time of the first communication task as $T_{i,0}$ and the end time of the last communication task as $T_{i,1}$. Then let $T_i$ be the smallest time so that all the communication tasks that begin at some time inside $[T_{i,0}, T_{i,0} + T_{i,1} - T_i]$ do not conflict with any of the communication tasks that end at some time inside $[T_i, T_{i,1}]$. Take the maximum over all nodes of the difference $T_i - T_{i,0}$, and denote it as $\Delta$. It must be true that $T(m\sum_k M_k) = T(\sum_k M_k)+(m-1)\Delta$ for any number of groups of packets $m$. This statement can be proven by induction on $m$.

First, consider the case when $m=2$. Since the same schedule is used cyclically, the delay between the starting time of the second schedule and the starting time of the first cycle is enough to determine the schedule used for two groups of data. Then, by definition, for any node $i$, if the second cycle begins at any time $T$ strictly earlier than $T_i - T_{i,0}$, $T+T_{i,0} < T_i$, and $T+(T_{i,0}+T_{i,1}-T_i) < T_{i,1}$. Therefore, the time interval $[T+T_{i,0},T+(T_{i,0}+T_{i,1}-T_i)]$ must overlap with the time interval $[T_i, T_{i,1}]$. By definition of $T_i$, such overlap must cause some tasks that begin at some time inside $[T_{i,0}, T_{i,0} + T_{i,1} - T_i]$ in the second schedule to conflict with some communication tasks that end at some time inside $[T_i, T_{i,1}]$ in the first schedule. And so, the second schedule must begin at some time later than or equal to $T_i$ for any node $i$; this means that the second schedule must begin at some time later than $\Delta$. Lastly, by the definition of $\Delta$, the second schedule can begin at $\Delta$ without causing any conflicts with the first schedule. So the overall finishing time for $m=2$ is just $T(\sum_k M_k)+\Delta$ as the second schedule begins at time $\Delta$, and the claim holds for $m=2$.

Second, assume the claim is true for $m$, then for $m+1$, by the same reason as the argument for $m=2$, the $m$-th schedule will start at $(m-1)\Delta$, and so the $m+1$-th schedule must not start strictly earlier than $(m-1)\Delta + \Delta = m\Delta$. Now, if the $m+1$-th cycle starts at $m\Delta$, by the same reason as the argument for $m=2$, it will not cause any conflict with the $m$-th schedule. And to see that it will also not cause any conflict with any earlier $m'$-th schedules with $m'<m$, just note that for any node $n$, $m\Delta > m'\Delta \geq (m'-1)\Delta + T_i$. So, by the definition of $\Delta$ and $T_i$, the $m$-th schedule will not cause any conflict with any of the earlier schedules. Thus, the claim holds for $m+1$ as well.

Finally, $T(M)= T(m\sum_k M_k) = T(\sum_k M_k)+(m-1)\Delta$. Setting $a=T(\sum_k M_k) - \Delta$ and $b=\Delta$ confirms the theorem, as $m=\frac{M}{\sum_k M_k}$.
\end{proof}

To further analyze the time profile and to better understand how message partitioning would affect the time profiles of different possible schedules, more structural assumptions are necessary. Particularly, denote $D=\max\limits_{(i,j)\in\vec{E}} (L_{i,j}B_{i,j})$ as the maximum latency-bandwidth product among all edges. Under the assumption that all packet sizes $M_k$ are far larger than $D$, it is possible to derive the optimal group of packet sizes $\sum_k M_k$ that will minimize the broadcast time. When the assumptions are not true, this will serve as an overestimation of the minimum broadcast time.

Now some notations are introduced. Let the time required to send a packet of size $n$ along a directed edge $(i,j)$ be $ L_{i,j} + \frac{n}{B_{i,j}}$, where $L_{i,j}$ is the latency and $B_{i,j}$ is the bandwidth of the edge. Under the assumption that $M_k \gg D \ \ \forall k$, let $L$ denote the minimal latency and $B$ denote the maximum bandwidth. For an arbitrary edge $(i,j)$ and packet index $k$ it is true that:

$$\frac{L_{i,j} + \frac{M_{k_1}}{B_{i,j}}}{L+\frac{M_{k_2}}{B}} = \frac{L_{i,j}B + \frac{M_{k_1}B}{B_{i,j}}}{LB+M_{k_2}} \simeq \frac{M_{k_1}B}{M_{k_2}B_{i,j}} = \frac{B}{B_{i,j}} \cdot \frac{M_{k_1}}{M_{k_2}}$$

This expression depends only on the relative size ratio of the packets $\frac{M_k}{\sum_k M_k}$, and is independent of the size of the packet groups $\sum_k M_k$. Therefore, no matter how the size $\sum_k M_k$ is changed, as long as the relative size ratios $\frac{M_k}{\sum_k M_k}$ remain the same, $L+\frac{\min\limits_k M_k}{B}$ can be the unit time, and a schedule can be made based on the time ratios. Setting $\tau_k = L + \frac{M_k}{B}$, the unit time can be denoted as $\min\limits_k \tau_k$. As long as the assumption is true, the same schedule will always remain a valid schedule, and its overall ratios:
\begin{align*}
\hat{a} &= \frac{T\left(\sum_{k=1}^K M_k\right)}{\min\limits_k \tau_k}, \\
\hat{b} &= \frac{\Delta}{\min\limits_k \tau_k}.
\end{align*}
will always remain the same, as all of the pairwise communication task time ratios also remain the same.

Now, consider the case where the total message of size $M$ is split into $m$ equally sized groups of packets, with size $\sum_k M_k$. Then consider a tree-based pipelined broadcast algorithm that has $K$ trees and a schedule with a time profile given by Equation \ref{time_prof_eq}. Using the factorization above, denote packet size-independent parameters $a=\hat{a}$ and $b=\hat{b}$. Because $\sum_k M_k = \frac{M}{m}$, the total broadcast time $T(M)$ can be written as:
\[
T(M) = (\hat{a} + \hat{b} m)(L + \frac{M}{mB}) = \hat{a}L + \frac{\hat{a}M}{mB} + m\hat{b}L + \frac{\hat{b}M}{B}.
\]
Optimizing $T(M)$ with respect to $M$ gives
\[
\frac{\partial T(M)}{\partial m} = - \frac{\hat{a}M}{B m^2} + \hat{b} L = 0,
\]
which yields the optimal number of packets and packet size
\begin{equation}
\label{m_opt}
m^\text{opt} = \sqrt{\frac{\hat{a}M}{\hat{b}LB}}, \frac{M}{m^\text{opt}} = \sqrt{\frac{\hat{b} L BM}{\hat{a}}}.
\end{equation}
These values must minimize the broadcast time, as $\frac{\partial ^2T(M)}{\partial M^2} < 0$. Substituting back, the corresponding optimal total broadcast time is
\begin{equation}
\label{t_opt}    
T^\text{opt} = \hat{a} L + \frac{\hat{b} M}{B} + 2 \sqrt{\frac{\hat{a}\hat{b}LM}{B}}.
\end{equation}

Denoting $\hat{B} = M/B$, this can be rewritten as
\[
T^\text{opt} = \hat{a} L + \hat{b} \hat{B} + 2 \sqrt{\hat{a}\hat{b}L\hat{B}},
\]
which allows comparison of different tree-based algorithms according to the constants $L$ and $\hat{B}$.  

This paper addresses certain difficulties in further analyzing how the performance of each of the arborescence solver and schedule solver would affect the time profile of the final pipelined algorithm constructed with these two solvers in the general case. It is always true that optimizing the actual broadcast time is the same as optimizing the factors $a$ and $b$ in Equation \ref{time_prof_eq}. However, in the most general setting, the influence of these factors on the scheduling and arborescence solvers does not admit an accurate objective function.

For $T(\sum_k M_k)$, the best general result without knowing the concrete rules that define the intersection graph $G_I$ is that $T(\sum_k M_k) \sim \Omega(\max\limits_k \ \tau_k \cdot \text{depth}(T_k))$, with $\text{depth}(T_k)$ measuring the depth of tree $k$. This is because a node can never send a packet before it receives the packet. This general result can be sharp when $G_I$ is unrealistically simple, such as each directed edge only intersecting with itself. Here, the bound provided is usually an underestimate that is much lower than the actual broadcast time for general $G_I$.

For $\frac{\Delta}{\sum_k M_k}$, a schedule must be solved heuristically. First, allow all communication tasks from all trees to be active simultaneously, then let the schedule solver construct a schedule that tries to minimize the finishing time. Then, record the best finishing time as $T^{\text{opt}}$. By definition of $\Delta$, $\Delta \sim \Omega(T^{\text{opt}})$ while $\sum_k M_k \sim O(K)$. However, under different assumptions of $G_I$, the difference between $T^{\text{opt}}$ can be very large with respect to the total number of edges in the arborescence.

So concrete knowledge about $G_I$ is necessary for more detailed estimates of the overall broadcast time. The results for each of the different solvers under concrete assumptions will be discussed in the following sections.

\subsection{The Schedule Problem}\label{schedule_solver}
Now, the scheduling problem will be addressed, assuming the arborescence problem was previously solved; thus, a collection of spanning trees is already known. As previously stated, if the finishing time of the schedule solver is $T^{\text{opt}}$ under the assumptions that all communication tasks must be scheduled together as tightly as possible, $T^{\text{opt}}$ bounds $\Delta$ in the time profile from below. The usual finishing time objective can be directly taken as the objective of the schedule problem. Note that after taking this objective, the schedule problem under the full-duplex constraint can be reduced to the classical open shop problem, and the schedule problem under the half-duplex constraint can be reduced to the file transfer problem. Thus, using these reductions, known results about the NP-hardness of this problem under general settings of arbitrary latency and bandwidth are first summarized, with known approximation algorithms.

Deriving an optimal schedule with arbitrary bandwidths and latencies, given a fixed spanning tree for each packet, is NP-complete under a half-duplex constraint, given as Theorem 1 in \cite{scheduling_file_transfers}, and NP-hard under a full-duplex constraint, as shown in Theorem 2 in \cite{short_shop_sch}. In the half-duplex setting, approximation algorithms that achieve suboptimal pipeline times within a factor of 3 are known, given by Corollary 10.1 in \cite{scheduling_file_transfers}. In the full-duplex setting, an approximation algorithm within a factor of 2 is known by Theorem 2.1 in \cite{approx_alg_shop_sch}, with the absolute best approximation factor proved as $\frac{5}{4}$ with no known algorithm, which is shown as Corollary 7 in \cite{short_shop_sch}. These approximation ratios are optimal under the respective general assumptions when spanning trees for each packet are fixed.

A true optimal solution can be achieved if all edges have the same bandwidth and latency, packets are the same size, and with the full-duplex constraint. In this case, a synchronous model can be considered, as all edges transmit equally sized packages at the same rate, so tasks are synchronized across all nodes and complete at the same time. A node that does not have a task at a certain time can simply wait for the next task and be synchronized. In this case, a set of instructions can be constructed that defines a cyclical communication protocol.

\begin{definition}
Under the assumptions of equal bandwidths and latencies, equally sized packets, and the full-duplex constraint, let $K$ directed spanning trees on a graph $\vec{G}$ be given. A \emph{pipeline} $\mathcal{P}$ is an ordered collection of $d$ edge sets
\[
\mathcal{P} = (E_1, E_2, \dots, E_d),
\]
such that within each set $E_i$, all directed edges can be activated simultaneously without conflict. In other words, no node is required to send or receive more than one packet at the same time. It is assumed that each set $E_i$ contains as many edges as possible subject to this constraint.
\end{definition}

\begin{theorem}
For a tree-based pipelined broadcast algorithm with $K$ directed trees, let $d$ be the maximum of the sum of the outgoing degrees across all trees. Then an optimal pipeline can be constructed with $d$ edge sets.
\label{pipe_const_thm}
\end{theorem}

\begin{proof}
From the underlying graph $G = (V, E)$, construct a bipartite multigraph $G^{*} = (V_1 \cup V_2, E^{*})$ as follows. For each node $v \in V$, create two copies: $v_1 \in V_1$ and $v_2 \in V_2$. Next, iterate over all $K$ directed trees. For each directed edge $(v_1,v_2)$ appearing in any tree, add an edge from $v_1 \in V_1$ to $v_2 \in V_2$ in $G^{*}$. Since edges may appear multiple times across different trees, $G^{*}$ is a multigraph. As all edges are between nodes within $V_1$ and $V_2$, with none within either set, $G^{*}$ is also bipartite. Thus, $G^{*}$ is a bipartite multigraph.

By construction, the degree of any node $v_1 \in V_1$ equals the total number of outgoing edges from $v_1$ across all trees, and hence is at most $d$. Similarly, all nodes in $G^{*}$ have degree at most $d$.

Any bipartite multigraph with maximum degree $d$ admits an edge coloring using exactly $d$ colors, which can be found in polynomial time \cite{10.1145/800133.804346}. Assign each color class to a set $E_i$ in a pipeline $\mathcal{P}$. Since each color class is a matching, no node has more than one outgoing edge in any set, and thus all edges in $E_i$ can be activated simultaneously. Since each color packs the maximum number of edges by definition, the matching is optimal. Therefore, $\mathcal{P}$ contains exactly $d$ edge sets. Because each outgoing edge at the same node must have a different color, any schedule of length $d$ must be optimal. 

\end{proof}

With the pipeline $\mathcal{P}$, which defines the ordered set of simultaneous active edges to send one group of packets, these edge sets can be simply repeated until all groups of packets are sent, so that all nodes contain the full message. Given a set of trees and the constraints of full-duplex, equal latency, equal bandwidth, and equal-sized packets, Theorem \ref{pipe_const_thm} demonstrates that the construction of a pipeline can be found through a bipartite multigraph edge coloring in polynomial time \cite{10.1145/800133.804346}. In this case, the exact optimum schedule can be constructed. Also note that, in the half-duplex case, if the graph $G$ is bipartite, the same theorem guarantees a strictly optimal schedule solution as well. And in case the graph $G$ is a general graph, though the edge coloring problem is NP-hard, proved in Section 1 of \cite{hd_exist}, a schedule can be found with no more than $d^*+1$ colors where $d^*$ is the true value of minimum number of colors required proved by Theorem 1.1 in \cite{hd_exist}, with an algorithm that completes in polynomial time given in Theorem 1.1 in \cite{hd_alg}.

The scheduling problem optimal solution can thus be solved under certain conditions once the arborescence problem is solved. However, the spanning trees must be the first construction to construct a pipelined broadcast algorithm, so an algorithm to solve the arborescence problem is required beforehand. 

\subsection{The Arborescence Problem}\label{tree_solver}
Now, the arborescence problem is addressed. First, as previously stated in \ref{time_profile}, both the factor $T(\sum_k M_k)$ and $\frac{\Delta}{\sum_k M_k}$ in the time profile depend on the solution of the arborescence problem. Thus, it is unavoidable to formulate the arborescence problem as a multi-criteria problem and optimize many criteria simultaneously. First, notice that aside from the general lower bound on $T(\sum_k M_k)$ derived in \ref{time_profile}, even with full-duplex assumptions, the outgoing edges in all trees from the root will conflict with each other, and all outgoing edges in any tree will also conflict with each other. Let $\text{deg}_\text{out}(T_k)$ denote the number of outgoing edges of a tree $T_k$, note that all such conflicts among all trees would introduce another term proportional to $\sum_k \text{deg}_\text{out}(T_k) \geq K +\max\limits_k \text{deg}_\text{out}(T_k)$ into the lower bound. Thus, from the lower bound on the term $a$ alone in the time profile, the following terms are in the objective:

\begin{equation}\label{term_a_lb}
    \max_k \left(\tau_k \cdot \text{depth}(T_k)\right) + \max_k \left(\tau_k \cdot \text{deg}_\text{out}(T_k)\right) + \sum\limits_k \tau_k,
\end{equation}
where $L$ is the minimal latency and $B$ is the maximal bandwidth, as denoted in Equation \ref{time_prof_eq}. In the half-duplex assumption, the situation is very similar: instead of $\text{deg}_\text{out}$, the total outgoing and incoming degrees must be considered in the same formula.

This objective is already a multi-criteria objective, and it is already too complex to optimize. Even with the assumption of equal efficiencies and a full-duplex constraint, when max degree and max depth among all trees appear together in the objective, the arborescence problem is NP-hard, as shown in Section 4 of \cite{Bicriteria}. In this case, to deal with the complexity in setting up a proper objective, approximation algorithms must be used, and then budgets must be imposed on terms in Equation \ref{term_a_lb}, and then finally, minimizing $\frac{\Delta}{\sum_k M_k}$ must be attempted.

Similar to the strategy in \ref{schedule_solver}, to set up a proper objective for the minimization of $\frac{\Delta}{\sum_k M_k}$, first, some lower bound of this term must be identified that can be computed for any feasible arborescence solution. To do this, a lower bound for $\Delta$ must be derived that only depends on the intersection graph $G_I$, tree $T_k$, and the packet size $M_k$ for all values of $k$. 

\begin{definition} 
For a set of edges $I$, it is called an intersecting edge set if any edge in the set intersects with all other edges.
\end{definition}

Allow all edges from all trees to be simultaneously active and identify all intersecting edge sets. For all edges in each intersecting edge set, compute the sum of completion times, and take the maximum value. This value bounds $\Delta$ from below by the definition of $\Delta$.

Denote this lower bound as $\Delta^*$, and then let the arborescence solver minimize $\frac{\Delta^*}{\sum_k M_k}$. Note that even for $\Delta^*$, directly minimizing it is NP-hard, as the sum of latency terms according to the intersection graph would make the objective function contain a term that tries to minimize the total outgoing degree of the $K$ many trees. This makes the problem NP-hard, as shown in section 4 of \cite{Bicriteria}. To derive approximation algorithms, if it is assumed that all $M_k$ are far larger than $D=\max\limits_{(i,j)\in\vec{E}} (L_{i,j}B_{i,j})$, further analysis can be done. Under this assumption, one observation is that:

\[
\frac{\Delta^*}{\sum_k M_k} = \frac{L+\frac{\sum_k M_k}{B}}{\sum_k M_k} \cdot \frac{\Delta^*}{L+\frac{\sum_k M_k}{B}}
\]
where the first term only depends on the size of packet groups $\sum_k M_k$. So, $L$ can be set as the minimal latency, $B$ as the maximal bandwidth, and $L+\frac{\sum_k M_k}{B}$ as the unit time. Then, for each directed edge $(i,j)$:
\[
\frac{L_{i,j} + \frac{M_{k}}{B_{i,j}}}{L+\frac{\sum_k M_k}{B}} \simeq \frac{BM_k}{B_{i,j}\sum_k M_k}.
\]
Thus, with some intersecting edge set $I$, the ratio $\frac{\Delta^*}{L+\frac{\sum_k M_k}{B}}$ only depends on fractions $\lambda_k = \frac{M_k}{\sum_k M_k}$ under previous assumptions. This can be shown by definition of $\Delta^*$:

\begin{align*}
\frac{\Delta^*}{L+\frac{\sum_k M_k}{B}} = & \frac{\max\limits_{I} \sum\limits_{(p,q) \in I, (p,q) \in T_l} L_{p,q} + \frac{M_l}{B_{p,q}}}{L+\frac{\sum_k M_k}{B}} \\
\simeq & B \cdot \max\limits_{I} \sum_{(p,q) \in I, (p,q) \in T_l}\frac{M_l}{B_{p,q}\sum_k M_k} \\ = & B \cdot \max\limits_{I} \sum_{(p,q) \in I, (p,q) \in T_l} \frac{\lambda_l}{B_{p,q}}.
\end{align*}

So this ratio may be minimized instead of raw time, as this ratio only depends on the relative packet size fractions $\lambda_k = \frac{M_k}{\sum_k M_k}$ and is independent of actual packet sizes. Using this new objective, a naive solution can be found for the weighted arborescence of spanning trees $T_k$ with weights $\lambda_k$ by minimizing $\max\limits_{I} \sum_{(p,q) \in I, (p,q) \in T_k} \frac{\lambda_{k}}{B_{p,q}}$ under the constraint that $\sum_k \lambda_k = 1$, where $I$ can be any intersecting edge set. However, this direct approach is a "min-max" problem and cannot be reduced by putting constraints on the weight $\lambda_k$'s due to the fact that some intersecting edge sets are fundamentally larger than others. As an alternative strategy, the quantity $\frac{\sum_k M_k}{\Delta^*}$ can be maximized. To do this, first note that under the assumption $M_k \gg D \ \ \forall k$, for any non-root node $j$ and its parent $i_{j,k}$ in tree $T_k$:

\[
\frac{\sum_k M_k}{\Delta^*} = \sum_k \frac{M_k}{L_{i_{j,k},j}+B_{i_{j,k},j}^{-1}\sum_k M_k} \cdot \frac{L_{i_{j,k},j}+B_{i_{j,k},j}^{-1}\sum_k M_k}{\Delta^*}.
\]
In addition to this
\[\frac{M_k}{L_{i_{j,k},j} + M_k(B_{i_{j,k},j})^{-1}} = \frac{B_{i_{j,k},j}M_k}{L_{i_{j,k},j}B_{i_{j,k},j} + M_k} \simeq B_{i_{j,k},j}
\] 
is independent of $M_k$. Then, under the same assumption, denote
\begin{align*}
O_{i_{j,k},j} = & \frac{L_{i_{j,k},j} + B_{i_{j,k},j}^{-1}M_k}{\Delta^*} \\
= & \frac{L_{i_{j,k},j} + B_{i_{j,k},j}^{-1}M_k}{\max\limits_{I} \sum\limits_{(p,q) \in I, (p,q) \in T_l} L_{p,q} + B_{p,q}^{-1}M_l} \\
\simeq & \frac{M_k}{B_{i_{j,k},j} \cdot \max\limits_{I} \sum\limits_{(p,q) \in I, (p,q) \in T_l} B_{p,q}^{-1}M_l} \\ 
= & \frac{\lambda_k}{B_{i_{j,k},j} \cdot \max\limits_{I} \sum\limits_{(p,q) \in I, (p,q) \in T_l} B_{p,q}^{-1}\lambda_l}
\end{align*}

Heuristically, $O_{i_{j,k},j}$ is the proportion of time needed for the task of node $i_{j,k}$ to send a message of size $M_{i_{j,k},j}$ to node $j$ over the maximal total time needed for all tasks in an intersecting edge set over all intersecting edge sets. But the most important property is that $O_{i_{j,k},j}$ is independent from the actual message size, and only depends on the relative ratios $\lambda_k$. If these $O_{i_{j,k},j}$ are used as raw variables rather than $\lambda_k$, our objective would become maximizing $\sum_k O_{i_{j,k},j}B_{i_{j,k},j}$, under constraints that $\sum_{k,j:(i_{j,k},j) \in I} O_{i_{j,k},j} \leq 1$ for all intersecting edge set $I$. 

Before further discussion of the problem, note that in the arborescence solver, it does not begin with an arborescence in the first place. Therefore, the domain for the formulation of the problem must be enlarged. This can be done by allowing $O_{i_{j,k},j}$ variables to be defined independently from a given arborescence of $K$ spanning trees. To build constraints on these variables, first start from a given arborescence and take its union, and for each directed edge $(i,j)$, define $O_{i,j} = \sum_{k: i_{j,k}=i} O_{i_{j,k},j}$. Heuristically, this $O_{i,j}$ is the total fraction of time spent on communication from nodes $i$ to $j$ across all trees. Then, for each node $j$, the quantity $\sum_k O_{i_{j,k},j}B_{i_{j,k},j}$ reduces to $\sum_i O_{i,j} B_{i,j}$, and this sum remains constant for all node $j$. Finally, the constraints $\sum_{k,j:(i_{j,k},j) \in I} O_{i_{j,k},j} \leq 1$ just reduce to $\sum_{(i,j) \in I} O_{i,j} \leq 1$ for any intersecting edge set $I$.

Now, these $O_{i,j}$ variables can be used to propose a linear programming (LP) problem to maximize $\sum_i O_{i,j} B_{i,j}$. The constraints, however, now come from 3 different origins. The first type is $\sum_{(i,j) \in I} O_{i,j} \leq 1$ as mentioned before. The second type comes from the fact that anything that cannot be generated from a weighted arborescence must be infeasible, particularly $O_{i, root} = 0$, $O_{i,j}B_{i,j} \leq \min(\sum_{k} O_{k, i}B_{k, i}, \sum_{k} O_{root,k}B_{root,k}) \ \ \forall i, j$, and $\sum_{j} O_{j, i}B_{j, i} = \mathcal{C} \ \ \forall i \neq root$, where $root$ is the index of the root node. The third type of constraints prevents obvious violations, particularly the constraints $O_{i,j}  = 0 \ \ \forall (i,j) \not\in \vec{E}$, and $0 \leq O_{i,j} \leq 1 \ \ \forall (i,j)$. And then under this formulation, any spanning tree $T$ is just a vector whose coefficient in front of each $O_{i,j}$ is $B_{i,j}$ if edge $(i,j) \in T$; or $0$ if edge $(i,j) \notin T$. Thus, an arborescence is just a vector basis, and the maximum value for an arborescence is just the maximum value of the LP restricted onto that basis. This turns the arborescence optimization problem into a basis search problem for the LP, whose complexity will be discussed after a discussion of the possible budgets for the $a$ term that adds all additional necessary complexities.

Unfortunately, if the budget is imposed on the max depth among all $T_k$ trees, the problem can be reduced to the bounded depth spanning tree problem by setting $K=1$ and setting all latencies and bandwidths equal, which is NP-hard as proved in section 1.3 in \cite{BDST}. So a budget must be set on the number of trees, with the hope that heuristic methods will not generate trees with large depths on general dense graphs. To the best of our knowledge, there is no known analysis of the joint optimization of the number of trees $K$ and matching the flow in a collection of trees to $\mathcal{C}$. As a suboptimal joint optimization, the number of trees can be arbitrarily fixed and incremented. It is important to note that $K$ must be small, since if the message size is not impractically large, the number of trees is guaranteed to increase execution time. In this case, a tree generation algorithm is proposed that tries to resolve this problem and gives bounds on optimality. This tree generation algorithm will construct $K$ trees that match as close as possible to the output of the LP.

Approximating an optimal incoming flow $\mathcal{C}$ using a collection of $K$ spanning trees with the same given root is known to be possible. In the general setting with arbitrary ${B_{i,j}}$, there exists a construction of $K$ spanning trees whose aggregated incoming flows achieve $\mathcal{C} - O\left(\sqrt{\frac{|V|\log |\vec{E}|}{K}}\right)$. This is given by Theorem 2.4 in \cite{fast_approx_alg}, where $|V|$ is due to the maximum congestion on a single tree, and $|\vec{E}|$ is due to the total number of constraints given by the intersection graph. 

Under the additional assumption of uniform latencies and bandwidths, as shown by Corollary 5.5.6 in \cite{iter_methods}, the approximation improves to $\mathcal{C} - O(\frac{d-1}{K + d -1})$, where $d$ is the maximum degree of the intersection graph. This result guarantees that $K$ spanning trees can be found with the maximum number of congestion no greater than $K+d-1$. With the additional assumption that all packets have the same size equal to $\hat{M}$, by definition $\Delta^* = (K+d-1)(L+\frac{\hat{M}}{B})$ under the full-duplex assumption, and $\Delta^* = (2K+d-1)(L+\frac{\hat{M}}{B})$ under the half-duplex assumption. Then the incoming efficiency for each package is given by $\frac{\hat{M}}{\Delta^*} \simeq \frac{B}{K+d-1}$ under the full-duplex assumption, or $\frac{\hat{M}}{\Delta^*} \simeq \frac{B}{2K+d-1}$ under the half-duplex assumption. Then the total incoming efficiency for this solution is $\frac{KB}{K+d-1}$ under the full-duplex assumption, and $\frac{KB}{2K+d-1}$ under the half-duplex assumption. But the maximum allowed total incoming efficiency is $B$ under the full-duplex assumption, or $\frac{B}{2}$ under the half-duplex assumption. Thus, taking the difference gives the desired results as:
\[
B - \frac{KB}{K+d-1} = B\left(\frac{d-1}{K+d-1}\right)
\]
for the full-duplex assumption, and
\[
\frac{B}{2}-\frac{KB}{2K+d-1}=B\left(\frac{d-1}{4K+2d-2}\right)
\]
for the half-duplex assumption.

Finally, in the regime where $K$  can be sufficiently large, the optimal incoming flow $\mathcal{C}$ can be achieved exactly with $O(|\vec{E}|^3)$ many trees even under arbitrary edge bandwidths, as proved in section VII.A from \cite{max_throughput}. Under the stronger assumption that all the edge bandwidths are rational multiples of each other, and thus all the $O_{i,j} B_{i,j}$ values are rational multiples of each other, the optimal incoming flow $\mathcal{C}$ can be achieved exactly with $O(|\vec{E}|)$ many trees, which is proved by Theorem 7.1 in \cite{edmonds}.

\subsection{Broadcast by Balanced Saturation}
\label{bbs}
Building on the results from previous sections, Broadcast by Balanced Saturation (BBS) provides a unified framework for constructing pipelined tree-based broadcast algorithms. BBS is a layered algorithm that fully addresses the broadcast problem. BBS first attempts to solve the LP in subsection \ref{tree_solver} under communication constraints specified by the intersection graph $G_I$. This LP outlines the best possible flow of data in the topology. Then, a suitable arborescence generation algorithm is chosen based on given assumptions. From this, a suitable scheduling algorithm is chosen, as discussed in subsection \ref{schedule_solver}. In each layer, BBS will select the optimal algorithm if assumptions permit, or a heuristic suboptimal algorithm based on the specific assumptions.

For further theoretical analysis, an example BBS algorithm under some common assumptions is proposed. In this case, bandwidth and latency are assumed to be equal for all connections with the full-duplex constraint. All connections capable of data transmission are contained as a set of directed edges $\vec{E}$. Furthermore, all packets are of equal size. This setting will reduce the broadcast problem and allow analysis and comparison with other broadcast algorithms. The constraints for the LP discussed in subsection \ref{tree_solver} can now be further explicitly written. The LP constraints, with $B_{i,j}=1 \ \forall i,j$ without loss of generality, are as follows:

\begin{enumerate}[label=\arabic*)]
    \item Graph Constraints:
    \[ 
    O_{i,j}  = 0 \quad \forall (i,j) \not\in \vec{E}
    \]
    \[
    0 \leq O_{i,j} \leq 1 \quad \forall (i,j) \in \vec{E}
    \]
    \[
    O_{i,root} = 0 \quad \forall i
    \] 
    These constraints enforce the topology conditions, and only allow valid directed edges to have occupancy rates between 0 and 1. \\
    
    \item Send and Receive Constraints:
    \[
    \sum_{j}O_{i,j} \leq 1 \quad \forall i
    \]
    \[
    \sum_{i}O_{i,j} \leq 1 \quad \forall j
    \]
    This ensures that a node cannot send to, nor receive from, multiple nodes simultaneously. \\

    \item Pair Constraint:
    \[
    O_{i,j} + O_{j,i} \leq 1 \quad \forall i,j
    \]
    This constraint ensures that no two simultaneous transmissions in both directions can occur over the same physical link. \\
    
    \item Forwarding Constraint: 
    \[
    O_{i,j} \leq \sum_{k} O_{k,i} \quad \forall i, j
    \]
    This enforces causality by ensuring that any node's outgoing rate is bounded by its total incoming rate.
    \\

    \item Root Forwarding Constraint:
    \[
    O_{i,j}  \leq \sum_k O_{root,k} \quad \forall i,j
    \]
    This constraint ensures any given edge cannot transmit data at a higher rate than the total outgoing rate of the root node. \\

    \item Incoming Flow Constraint:
    \[
    \sum_{j} O_{j,i} = \mathcal{C} \ \ \forall i \neq root
    \]
    This ensures that every non-root node receives data at the same rate $\mathcal{C}$, defining the broadcast throughput of the system.
\end{enumerate}

As before, the LP optimizes for $\max(\mathcal{C})$. The existence of a solution is guaranteed; solutions to such linear optimization problems generally are not unique. In practice, a particular optimizer could be selected, often by imposing additional criteria such as symmetry of the topology. From the solved LP, a fixed number of trees is generated following subsection \ref{tree_solver}, chosen based on whether the number of trees must be small or can be large, depending on message size. Once these trees are defined, a suitable scheduling algorithm is chosen following subsection \ref{schedule_solver}.

For a given topology, the regime of message sizes can be identified directly from the ratio of $\frac{M}{D}$, where $M$ is the message size and $D=\max\limits_{(i,j)\in\vec{E}} (L_{i,j}B_{i,j})$, as shown in subsection \ref{time_profile}. From this, the appropriate broadcasting algorithm can be selected. In the regime of small message sizes, such as only having 1 packet of data, the $a$ term in Equation \ref{time_prof_eq} dominates the $b$ term. Here, the number of trees and their depth are most important to determine broadcast time. The maximum degree times the maximum depth is always greater than or equal to 2 times the diameter $\vec{G}$. In this case, BBS will generate a single shallow tree guided by the LP. In the regime of medium message sizes, neither term $a$ nor $b$ dominates the other. In this case, there are multiple but a small number of packets, and BBS resembles a heuristic algorithm where the number of trees is set to a small number $K$ and may be incrementally improved. In the LP, the total incoming flow $\mathcal{C}$ is shown to be close to optimal; however, with no restrictions on the depth of trees. In the regime of large message sizes, where there are a large number of packets, the $b$ term dominates the $a$ term. In this case, BBS falls back to solving the LP, but with the trees achieving the optimal incoming flow $\mathcal{C}$. Here, the number of trees $K$ is allowed to be large. In this case, the proposed BBS algorithm achieves asymptotic optimality.

\begin{figure}[H]
\centering
\includegraphics[width=1\linewidth]{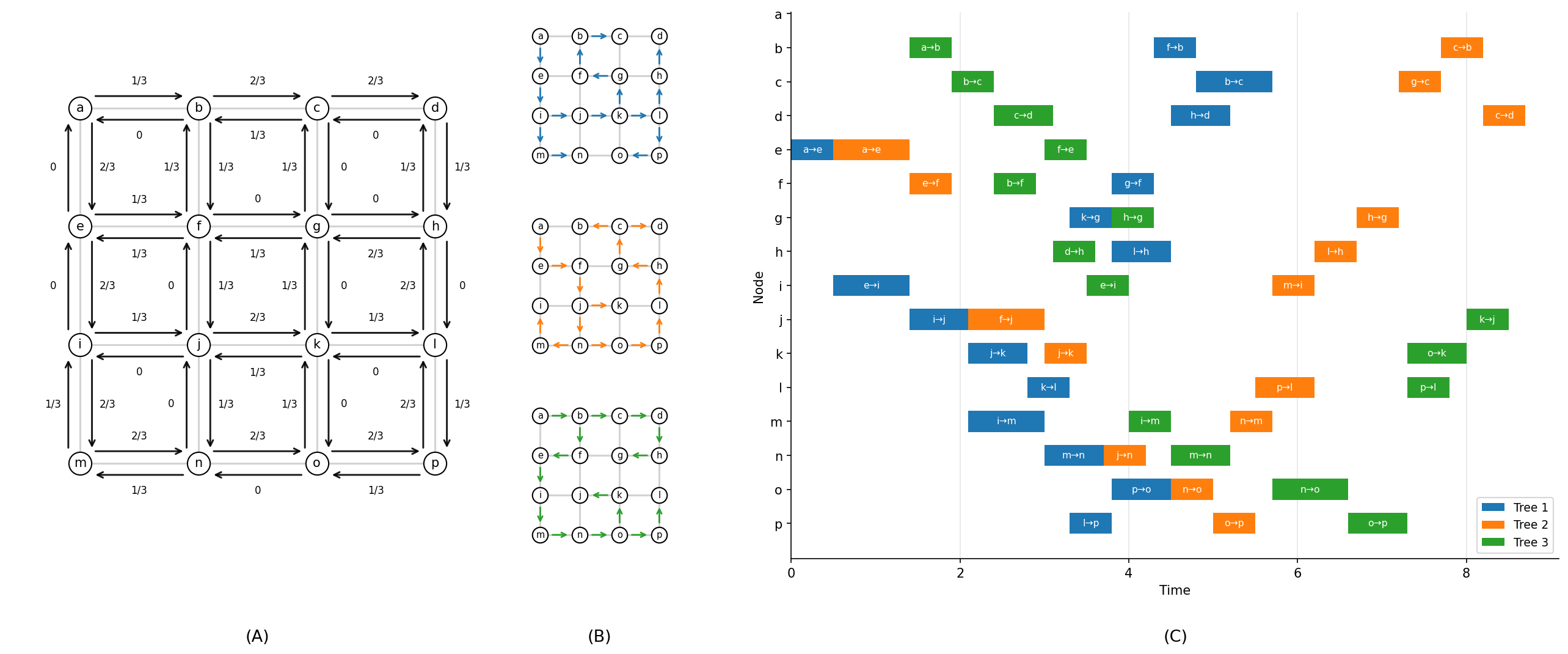}
\caption{Example construction of a BBS algorithm on a $4\times4$ 2D mesh topology, rooted at node $a$. (A) Solution of the LP formulation restricted to $K=3$ trees. (B) Tree construction from the LP solution. (C) Communication schedule constructed from the arborescences.}
\label{BBS}
\end{figure}

Finally, figure \ref{BBS} illustrates the complete construction of a BBS algorithm on a $4\times4$ 2D mesh topology, rooted at node $a$. Subfigure (A) shows the solution of the LP formulation restricted to $K=3$ trees under the constraints above. Each directed edge is labeled with its corresponding $O_{i,j}$ value, representing the total fraction of time for communication from node $i$ to node $j$ across all trees. From the LP solution, trees are extracted as shown in subfigure (B). These trees provide an explicit routing structure that is then used for scheduling. The resulting communication schedule is shown in subfigure (C), where communication tasks are scheduled while ensuring conflict-free execution. Each rectangle represents a communication task from a sending node to a receiving node and is plotted for the receiving node. To demonstrate the ability to handle nonuniform communication costs, each communication task is assigned a random duration between 5 and 10 time units. This emulates variability that may arise in practical systems, such as heterogeneous bandwidth and latency, or time-varying network conditions.

\section{Broadcast Simulations}
\subsection{Implementation Details}
All simulations are conducted using SimGrid \cite{simgrid}, a widely used simulation framework for distributed applications on large-scale systems, which is well-suited for evaluating broadcast algorithms. SimGrid allows for both realistic and reproducible simulations. Simulated MPI (SMPI) \cite{SMPI} within SimGrid will be used for programming the actual implementations.

In these simulations, the topologies 2D Mesh, Butterfly, Dragonfly, and Fat-Tree are used, where 2D Mesh and Butterfly are treated as nonhierarchical topologies, while Dragonfly and Fat-Tree are treated as hierarchical topologies. Each class of topologies is constructed with a number of nodes $N$ equal to 128, 256, 512, and 1024. The message sizes for the simulations are 64 and 256 KB, and 1, 4, 16, 64, and 128 MB. The 2D Mesh topology was chosen in this work to use InfiniBand NDR 400 links with a bandwidth of 50 GBps and a latency of 100 ns, with topology dimensions $8\times16$, $16\times16$, $8\times32$, and $32\times32$. The Butterfly topology uses InfiniBand EDR links with a bandwidth of 12.5 GBps and a latency of 100 ns \cite{flat_butterfly}. The Fat-Tree topology uses InfiniBand EDR links with a bandwidth of 12.5 GBps and a latency of 100 ns on all links \cite{fat_tree}. The Dragonfly topology uses Cray Aries links with a bandwidth of 5.25 GBps on all link types, but with varying latency: 100 ns for intra-chassis links, 200 ns for inter-chassis links, and 400 ns for inter-group links \cite{dragonfly}.

Regardless of topology, the BBS algorithm is constructed in three stages: defining the intersection graph $G_I$, constructing an arborescence subject to the constraints imposed by $G_I$, and constructing a schedule from the resulting arborescence. For the 2D Mesh and Butterfly topologies, all edges correspond directly to physical links between nodes in this implementation. Here, two directed edges are considered to intersect if they share a start or end node, or correspond to opposite directions of the same undirected edge. Edge hopping is intentionally prohibited in these topologies to preserve the influence of the underlying graph structure, rather than masking it through arbitrary path selections. Once $G_I$ is defined, the LP shown in subsection \ref{bbs} can be constructed and solved, after which the incremental arborescence algorithm from Section 5.5 of \cite{iter_methods} and the greedy scheduling algorithm from \cite{scheduling_file_transfers} are applied. Since all bandwidths, latencies, and packet sizes are uniform in this setting, a truly optimal schedule can theoretically be obtained through bipartite multigraph edge coloring, as discussed in subsection \ref{schedule_solver}. However, implementing such a solution in MPI is impractical, so the greedy algorithm is used instead.

In the BBS implementation for the Dragonfly and Fat-Tree topologies, communication occurs over virtual edges realized as paths of physical links, making the definition of $G_I$ dependent on both the virtual edge and its chosen physical realization. Since a single virtual edge may correspond to multiple physical paths, this work simplifies the problem by greedily fixing one realization for each virtual edge while minimizing the use of inter-group physical links. Two virtual directed edges are then considered to intersect whenever any physical edges in each of their underlying physical paths intersect according to the rule previously discussed. Using this definition of $G_I$, the arborescence and schedule constructions follow the same algorithms as before, except that the LP variables correspond to virtual rather than physical edges.

The baseline algorithms evaluated in this work are as follows. Global Links First (GLF) is a topology-aware broadcast algorithm designed for hierarchical networks. Although originally developed for Dragonfly topologies, it can also be applied to Fat-Tree topology. The implementation used in our experiments is taken from \cite{GLF}, which was originally introduced in \cite{GLF2}. For Dragonfly topologies, GLF proceeds from the coarsest to the finest hierarchy level as described in \cite{GLF}, selecting one representative per sub-group and broadcasting to them via a binomial tree. Each recipient then becomes the message holder for the next finer level, until a final intra-router broadcast completes delivery. For nonhierarchical topologies, GLF performs a BFS search to generate virtual ranks for each node and then constructs a binomial tree based on these virtual ranks to do the broadcast. MPI\_Bcast, the standard MPI broadcast function, is already implemented in SMPI within SimGrid, which will dynamically select an algorithm. Scatter and Recursive Doubling Allgather (SRDA), as well as Pipelined Chain Broadcast (Pipeline), are both implemented based on \cite{MPICH}. Finally, Binomial Negabinary Trees (Bine) is a special tree-based broadcast algorithm, with implementation details provided in \cite{Bine}.

All simulations will follow the full-duplex assumption with equally sized packets. In addition, the broadcast execution time will be measured excluding the time to construct the communication schedule. This is justified because once given a topology with communication conditions, the schedule can be constructed once, stored, and reused across executions, even with different message sizes. In simulations, a broadcast is completed over every node as the root node, and summary statistics are given for comparison.

\subsection{Results and Discussions}
\begin{figure}[H]
\centering
\includegraphics[width=1\linewidth]{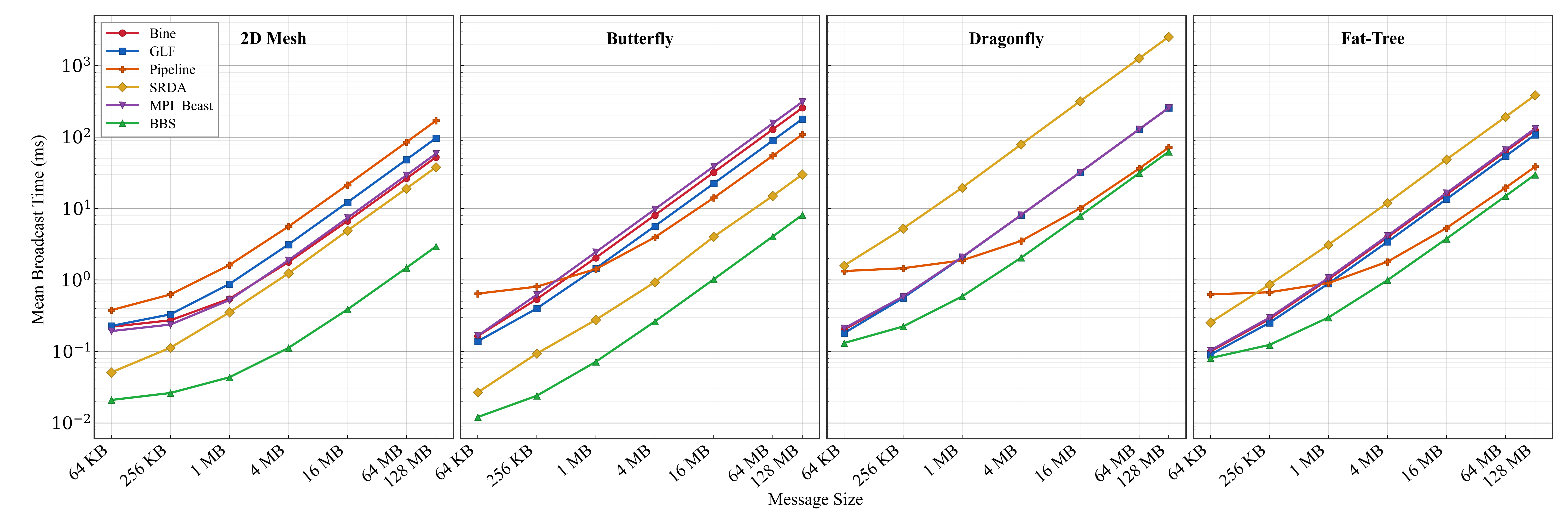}
\caption{$T_{\text{avg}}$ across various message sizes on 512-node topologies.}
\label{time_512}
\end{figure}

Tables \ref{tab:results_N128_nonrouter}-\ref{tab:results_N1024_router} in Appendix \ref{time_tables} present performance comparisons of the broadcast algorithms, with each table showcasing topology sizes of 128, 256, 512, and 1024 nodes, respectively. Within each table, summary statistics of the broadcast times are reported, including the average $T_{\text{avg}}$, maximum $T_{\text{max}}$, minimum $T_{\text{min}}$, and standard deviation $\sigma$. For each message size, the minimum $T_{\text{avg}}$ is shown in bold. 
Figure \ref{time_512} presents a log-log plot of $T_{\text{avg}}$ for all evaluated message sizes on topologies with 512 nodes. The log-log plots for all other topology sizes are provided as Figures \ref{time_128}, \ref{time_256}, and
\ref{time_1024} in Appendix \ref{time_figs}. Figure \ref{time_512_linear} presents a linear-scale plot of $T_{\text{avg}}$ for all topologies with 512 nodes. In this plot, a cut-off time is necessary for visual clarity, as the SRDA broadcast time is extremely long in the Dragonfly topology.

\begin{figure}[H]
\centering
\includegraphics[width=\linewidth]{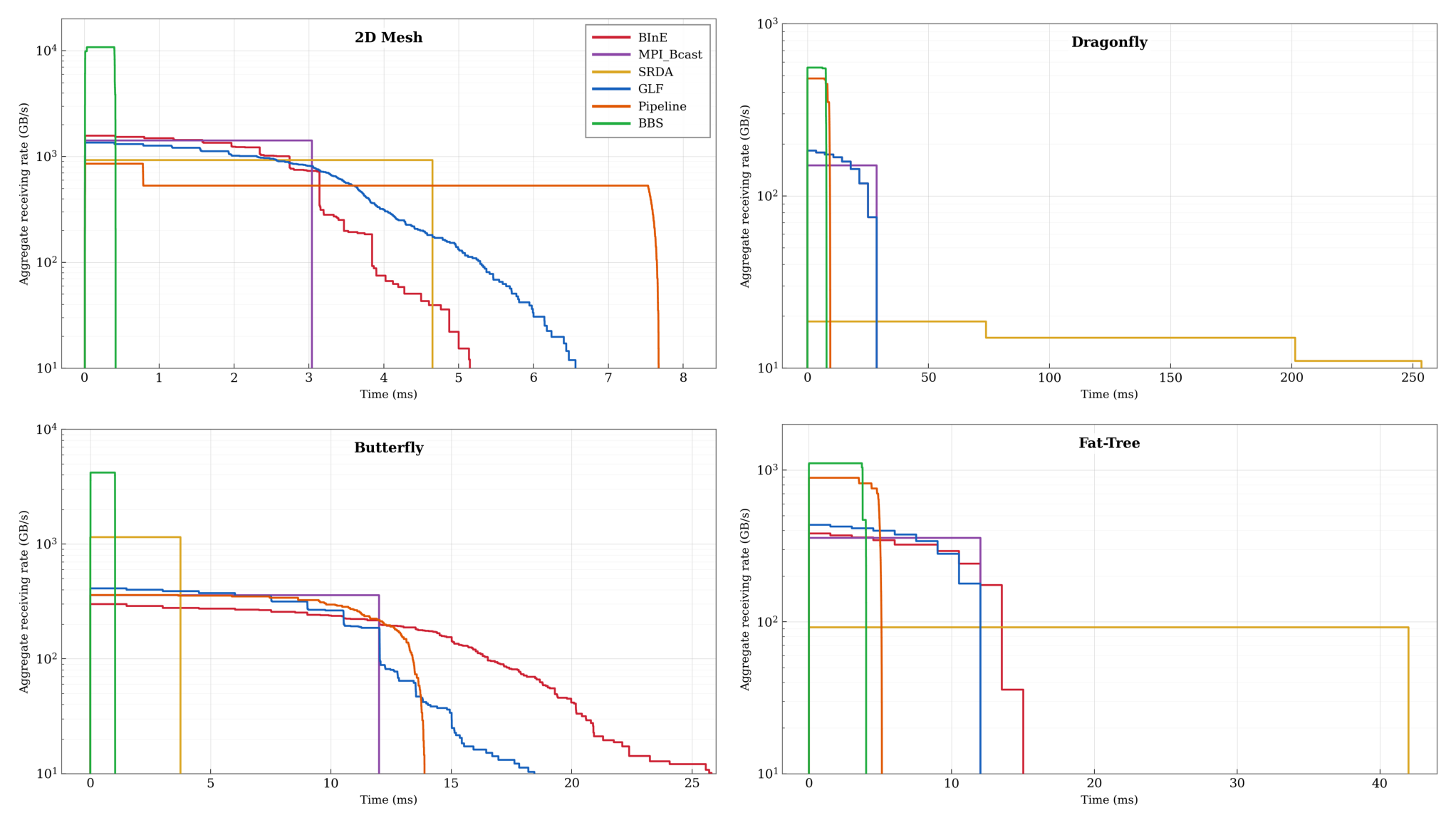}
\caption{Aggregated receiving rate of data over time for broadcasting 16 MB of data on topologies with 256 nodes.}
\label{rates}
\end{figure}

To better understand the dynamic behavior of broadcasts with different algorithms, aggregated data receiving rates over time were derived from SimGrid execution traces to approximate the instantaneous system-wide throughput. With a message size of 16 MB, Figure \ref{rates} presents the throughput rate figures over time for all topologies with 256 nodes. The overall shape of the curves provides meaningful insights into pipeline efficiency and bandwidth utilization.

Across all simulation results, regardless of topology and message size, the BBS algorithm outperforms all baseline algorithms by a substantial margin. Some algorithms tend to perform well on certain topologies, such as Pipeline on hierarchical networks and SRDA on nonhierarchical networks. In addition, certain algorithms, most notably Pipeline, tend to perform poorly with low message sizes and improve with larger sizes. In contrast, BBS is able to perform extremely well, regardless of message size and topology conditions, due to the robustness of the general algorithm capable of adaptation.

A key observation from the rate plots is that BBS achieves both a higher peak receiving rate and sustained rates over time compared to all baseline methods. In particular, while the baseline algorithms exhibit staggered and tapered receiving rates, indicative of partial or uneven network utilization, BBS maintains a high throughput until completion. This directly explains the consistently lower broadcast times observed across all experiments.

As predicted by Equation \ref{time_prof_eq}, all algorithms exhibit broadcast times that scale approximately linearly with message size, which can be seen in Figure \ref{time_512_linear}. This validates the practicality and predictive power of the time profile, which enables broadcast time to be approximated once the parameters $a$ and $b$ are known, along with the optimal parameters from Equation \ref{m_opt} and the optimal broadcast time from Equation \ref{t_opt}.

\begin{figure}[H]
\centering
\includegraphics[width=1\linewidth]{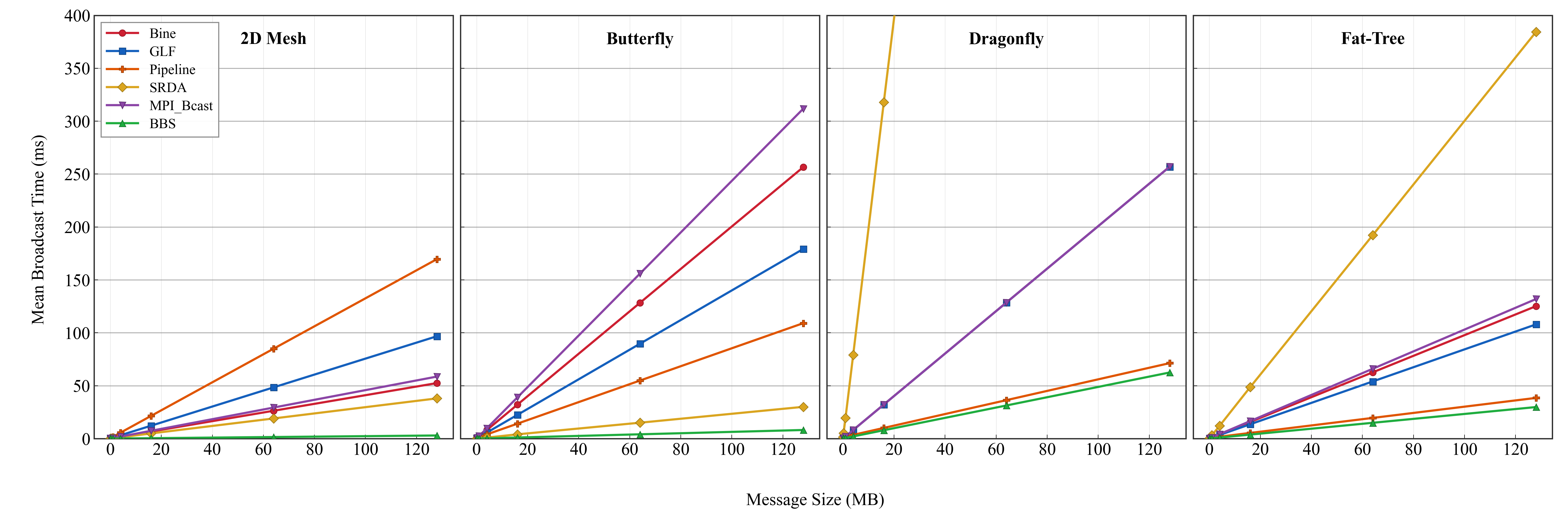}
\caption{$T_{\text{avg}}$ across various message sizes on 512-node topologies. For analysis and readability, a linear scale is used, and time is cut off at 400 ms.}
\label{time_512_linear}
\end{figure}

A notable and somewhat surprising result is that, in the Dragonfly and Fat-Tree topologies, only $K=2$ trees are required to achieve the maximal value of $\frac{\sum_k M_k}{\Delta}=\frac{B}{2}$ where $B$ is the uniform bandwidth in the time profile given in Section \ref{tree_solver}. The $\frac{1}{2}$ factor comes from the fact that all nodes only have one edge connecting to a router, and thus all incoming and outgoing communications at all nodes will conflict with each other due to the edge contention constraint. This suggests that these topologies admit highly complementary spanning tree pairs, likely due to their high symmetry and strong global connectivity. Consequently, once such a pair of trees is identified, it can be reused for all sufficiently large message sizes, eliminating the need for further optimization. For smaller messages, where pipeline effects are not significant, a single tree strategy is sufficient.

Overall, these results demonstrate that BBS not only outperforms existing broadcast algorithms in terms of raw performance in this setting, but also provides a unified framework that adapts effectively to diverse network structures and communication protocols.

\section{Conclusions and Future Work}
This paper presents BBS, a general class of tree-based pipelined broadcast algorithms that can be applied to a given network topology with specified communication protocols. BBS addresses key challenges in broadcast designs, including spanning tree construction and task scheduling, in both general and specific settings.

Under certain conditions, simulation results using SimGrid demonstrate that the BBS algorithm outperforms other broadcasting algorithms across a variety of topologies and message sizes. In particular, it maintains a strong performance advantage for increasing message sizes, reflecting its ability to effectively utilize bandwidth through pipelining. Moreover, the algorithm exhibits strong robustness with respect to network topology. While some baseline algorithms perform well with certain topologies due to structural assumptions, BBS maintains high performance across all tested topologies. This highlights the generalization capabilities of the algorithm and its ability to compete with and outperform topology-specific algorithms.

Several directions for future work remain. First, implementing BBS on real-world systems would enable validation of its performance with real-time experiments. Second, the framework can be extended to incorporate more general communication models, including relaxed or modified occupancy constraints, multicasting, or other advanced communication primitives. Third, the approach can be adapted to other collective communication operations, such as all-gather or multi-source broadcast. Finally, BBS may serve as a tool for network topology analysis and design, enabling the identification of structures that admit efficient broadcast schedules under constraints on nodes, degree, and connectivity.

\backmatter

\section*{Declarations}

\noindent\textbf{Author contributions} \
HL developed the theoretical concepts and components, with support from JH and BT. HL and JH performed the implementation, with assistance from BT and DH. BT composed the manuscript with support from HL, JH, and NT. HL, JH, BT, and NT contributed to the overall design and direction of this project. YD provided guidance and final revisions. All authors approved the final version of this publication. \\

\noindent\textbf{Code availability} \
The code for the different algorithms and simulations can be found on \href{https://github.com/Js-Hwang1/A-Bcast-Model-for-Given-Topologies.git}{GitHub}.\\

\noindent\textbf{Conflict of interest} \
The authors declare no conflicts of interest.




\bibliography{references}

@inproceedings{top_aware_infini,
author = {Subramoni, Hari and Kandalla, Krishna and Vienne, Jérôme and Sur, Sayantan and Barth, William and Tomko, Karen and Mclay, Robert and Schulz, Karl and Panda, D.K.},
year = {2011},
month = {09},
pages = {317-325},
title = {Design and Evaluation of Network Topology-/Speed- Aware Broadcast Algorithms for InfiniBand Clusters},
journal = {Proceedings - IEEE International Conference on Cluster Computing, ICCC},
doi = {10.1109/CLUSTER.2011.43}
}

@inproceedings{MPI_bluegene,
author = {Alm\'{a}si, George and Heidelberger, Philip and Archer, Charles J. and Martorell, Xavier and Erway, C. Chris and Moreira, Jos\'{e} E. and Steinmacher-Burow, B. and Zheng, Yili},
title = {Optimization of MPI collective communication on BlueGene/L systems},
year = {2005},
isbn = {1595931678},
publisher = {Association for Computing Machinery},
address = {New York, NY, USA},
url = {https://doi.org/10.1145/1088149.1088183},
doi = {10.1145/1088149.1088183},
booktitle = {Proceedings of the 19th Annual International Conference on Supercomputing},
pages = {253–262},
numpages = {10},
location = {Cambridge, Massachusetts},
series = {ICS '05}
}

@ARTICLE{novel_pipelined,
  author={Wu, Chi-Jen and Ku, Chin-Fu and Ho, Jan-Ming and Chen, Ming-Syan},
  journal={IEEE Transactions on Knowledge and Data Engineering}, 
  title={A Novel Pipeline Approach for Efficient Big Data Broadcasting}, 
  year={2016},
  volume={28},
  number={1},
  pages={17-28},
  doi={10.1109/TKDE.2015.2468714}}

@ARTICLE{max_throughput,
  author={Zongpeng Li and Baochun Li and Lau, L.C.},
  journal={IEEE Transactions on Information Theory}, 
  title={On achieving maximum multicast throughput in undirected networks}, 
  year={2006},
  volume={52},
  number={6},
  pages={2467-2485},
  doi={10.1109/TIT.2006.874515}}

@article{twotree,
title = {Two-tree algorithms for full bandwidth broadcast, reduction and scan},
journal = {Parallel Computing},
volume = {35},
number = {12},
pages = {581-594},
year = {2009},
note = {Selected papers from the 14th European PVM/MPI Users Group Meeting},
issn = {0167-8191},
doi = {https://doi.org/10.1016/j.parco.2009.09.001},
url = {https://www.sciencedirect.com/science/article/pii/S0167819109000957},
author = {Peter Sanders and Jochen Speck and Jesper Larsson Träff}
}

@INPROCEEDINGS{pipeline_bcast,
  author={Beaumont, O. and Legrand, A. and Marchal, L. and Robert, Y.},
  booktitle={18th International Parallel and Distributed Processing Symposium, 2004. Proceedings.}, 
  title={Pipelining broadcasts on heterogeneous platforms}, 
  year={2004},
  volume={},
  number={},
  pages={19-},
  doi={10.1109/IPDPS.2004.1302926}}

@article{compare_old_super,
title = {A performance comparison of current HPC systems: Blue Gene/Q, Cray XE6 and InfiniBand systems},
journal = {Future Generation Computer Systems},
volume = {30},
pages = {291-304},
year = {2014},
note = {Special Issue on Extreme Scale Parallel Architectures and Systems, Cryptography in Cloud Computing and Recent Advances in Parallel and Distributed Systems, ICPADS 2012 Selected Papers},
issn = {0167-739X},
doi = {https://doi.org/10.1016/j.future.2013.06.019},
url = {https://www.sciencedirect.com/science/article/pii/S0167739X13001337},
author = {Darren J. Kerbyson and Kevin J. Barker and Abhinav Vishnu and Adolfy Hoisie}
}

@ARTICLE{blue_gene,
  author={Adiga, N. R. and Blumrich, M. A. and Chen, D. and Coteus, P. and Gara, A. and Giampapa, M. E. and Heidelberger, P. and Singh, S. and Steinmacher-Burow, B. D. and Takken, T. and Tsao, M. and Vranas, P.},
  journal={IBM Journal of Research and Development}, 
  title={Blue Gene/L torus interconnection network}, 
  year={2005},
  volume={49},
  number={2.3},
  pages={265-276},
  doi={10.1147/rd.492.0265}}

@article{ring_torus_hyper,
title = {Ring, torus and hypercube architectures/algorithms for parallel computing},
journal = {Parallel Computing},
volume = {25},
number = {13},
pages = {1877-1906},
year = {1999},
issn = {0167-8191},
doi = {https://doi.org/10.1016/S0167-8191(99)00069-1},
url = {https://www.sciencedirect.com/science/article/pii/S0167819199000691},
author = {S. Lakshmivarahan and Sudarshan K. Dhall}
}

@article{hd_exist,
author = {Chen, Guantao and Jing, Guangming and Zang, Wenan},
year = {2025},
month = {09},
pages = {},
title = {Proof of the Goldberg–Seymour conjecture on edge–colorings of multigraphs},
volume = {50},
journal = {Journal of Combinatorial Optimization},
doi = {10.1007/s10878-025-01348-6}
}

@unknown{hd_alg,
author = {Chen, Guantao and Hao, Yanli and Yu, Xingxing and Zang, Wenan},
year = {2024},
month = {07},
pages = {},
title = {A short proof of the Goldberg-Seymour conjecture},
doi = {10.48550/arXiv.2407.09403}
}

@article{edmonds,
title = {Testing membership in matroid polyhedra},
journal = {Journal of Combinatorial Theory, Series B},
volume = {36},
number = {2},
pages = {161-188},
year = {1984},
issn = {0095-8956},
doi = {https://doi.org/10.1016/0095-8956(84)90023-6},
url = {https://www.sciencedirect.com/science/article/pii/0095895684900236},
author = {William H Cunningham}
}

@inproceedings{dragonfly,
author = {Kim, John and Dally, Wiliam J. and Scott, Steve and Abts, Dennis},
title = {Technology-Driven, Highly-Scalable Dragonfly Topology},
year = {2008},
isbn = {9780769531748},
publisher = {IEEE Computer Society},
address = {USA},
url = {https://doi.org/10.1109/ISCA.2008.19},
doi = {10.1109/ISCA.2008.19},
booktitle = {Proceedings of the 35th Annual International Symposium on Computer Architecture},
pages = {77–88},
numpages = {12},
series = {ISCA '08}
}

@article{fat_tree,
author = {Leiserson, Charles E.},
title = {Fat-trees:  universal networks for hardware-efficient supercomputing},
year = {1985},
issue_date = {Oct. 1985},
publisher = {IEEE Computer Society},
address = {USA},
volume = {34},
number = {10},
issn = {0018-9340},
journal = {IEEE Trans. Comput.},
month = oct,
pages = {892–901},
numpages = {10}
}

@article{SMPI,
author = {Degomme, Augustin and Legrand, Arnaud and Markomanolis, George S. and Quinson, Martin and Stillwell, Mark and Suter, Fr\'{e}d\'{e}ric},
title = {Simulating MPI Applications: The SMPI Approach},
year = {2017},
issue_date = {Aug. 2017},
publisher = {IEEE Press},
volume = {28},
number = {8},
issn = {1045-9219},
url = {https://doi.org/10.1109/TPDS.2017.2669305},
doi = {10.1109/TPDS.2017.2669305},
journal = {IEEE Trans. Parallel Distrib. Syst.},
month = aug,
pages = {2387–2400},
numpages = {14}
}

@article{approx_alg_shop_sch,
title = {Approximation algorithms for the multiprocessor open shop scheduling problem},
journal = {Operations Research Letters},
volume = {24},
number = {4},
pages = {157-163},
year = {1999},
issn = {0167-6377},
doi = {https://doi.org/10.1016/S0167-6377(99)00005-X},
url = {https://www.sciencedirect.com/science/article/pii/S016763779900005X},
author = {Petra Schuurman and Gerhard J. Woeginger}
}

@article{short_shop_sch,
author = {Williamson, D. P. and Hall, L. A. and Hoogeveen, J. A. and Hurkens, C. A. J. and Lenstra, J. K. and Sevast'janov, S. V. and Shmoys, D. B.},
title = {Short Shop Schedules},
year = {1997},
issue_date = {April 1997},
publisher = {INFORMS},
address = {Linthicum, MD, USA},
volume = {45},
number = {2},
issn = {0030-364X},
url = {https://doi.org/10.1287/opre.45.2.288},
doi = {10.1287/opre.45.2.288},
journal = {Oper. Res.},
month = apr,
pages = {288–294},
numpages = {7}
}

@article{fast_approx_alg,
author = {Plotkin, Serge A. and Shmoys, David B. and Tardos, \'{E}va},
title = {Fast Approximation Algorithms for Fractional Packing and Covering Problems},
journal = {Mathematics of Operations Research},
volume = {20},
number = {2},
pages = {257-301},
year = {1995},
doi = {10.1287/moor.20.2.257},
URL = {https://doi.org/10.1287/moor.20.2.257},
eprint = {https://doi.org/10.1287/moor.20.2.257}
}

@book{iter_methods,
  title     = {Iterative Methods in Combinatorial Optimization},
  author    = {Lau, Lap-Chi and Ravi, R. and Singh, Mohit},
  year      = {2011},
  address   = {Cambridge, UK},
  publisher = {Cambridge University Press}
}

@article{Bicriteria,
title = {Bicriteria Network Design Problems},
journal = {Journal of Algorithms},
volume = {28},
number = {1},
pages = {142-171},
year = {1998},
issn = {0196-6774},
doi = {https://doi.org/10.1006/jagm.1998.0930},
url = {https://www.sciencedirect.com/science/article/pii/S0196677498909300},
author = {Madhav V Marathe and R Ravi and Ravi Sundaram and S.S Ravi and Daniel J Rosenkrantz and Harry B Hunt}
}

@inproceedings{scheduling_file_transfers,
author = {Coffman, E. G. and Garey, M. R. and Johnson, D. S. and LaPaugh, A. S.},
title = {Scheduling file transfers in a distributed network},
year = {1983},
isbn = {0897911105},
publisher = {Association for Computing Machinery},
address = {New York, NY, USA},
url = {https://doi.org/10.1145/800221.806726},
doi = {10.1145/800221.806726},
booktitle = {Proceedings of the Second Annual ACM Symposium on Principles of Distributed Computing},
pages = {254–266},
numpages = {13},
location = {Montreal, Quebec, Canada},
series = {PODC '83}
}

@inproceedings{bine,
author = {De Sensi, Daniele and Pasqualoni, Saverio and Piarulli, Lorenzo and Bonato, Tommaso and Ba, Seydou and Turisini, Matteo and Domke, Jens and Hoefler, Torsten},
title = {Bine Trees: Enhancing Collective Operations by Optimizing Communication Locality},
year = {2025},
isbn = {9798400714665},
publisher = {Association for Computing Machinery},
address = {New York, NY, USA},
url = {https://doi.org/10.1145/3712285.3759835},
doi = {10.1145/3712285.3759835},
booktitle = {Proceedings of the International Conference for High Performance Computing, Networking, Storage and Analysis},
pages = {1901–1916},
numpages = {16},
series = {SC '25}
}

@INPROCEEDINGS{fat_tree_alg,
  author={Kumar, Sameer and Sharkawi, Sameh S. and Jan, K. A. Nysal},
  booktitle={2016 IEEE International Parallel and Distributed Processing Symposium (IPDPS)}, 
  title={Optimization and Analysis of MPI Collective Communication on Fat-Tree Networks}, 
  year={2016},
  volume={},
  number={},
  pages={1031-1040},
  doi={10.1109/IPDPS.2016.85}}

@inproceedings{GLF,
  title={Evaluation of topology-aware broadcast algorithms for dragonfly networks},
  author={Dorier, Matthieu and Mubarak, Misbah and Ross, Rob and Li, Jianping Kelvin and Carothers, Christopher D and Ma, Kwa-Liu},
  booktitle={2016 IEEE International Conference on Cluster Computing (CLUSTER)},
  pages={40--49},
  year={2016},
  organization={IEEE}
}

@article{GLF2,
author = {Xiang, Dong and Liu, Xiaowei},
year = {2015},
month = {01},
pages = {1-1},
title = {Deadlock-Free Broadcast Routing in Dragonfly Networks without Virtual Channels},
volume = {27},
journal = {IEEE Transactions on Parallel and Distributed Systems},
doi = {10.1109/TPDS.2015.2503746}
}

@article{simgrid,
  title = {{Lowering entry barriers to developing custom simulators of distributed applications and platforms with SimGrid}},
  journal = {Parallel Computing},
  volume = {123},
  pages = {103-125},
  year = {2025},
  issn = {0167-8191},
  doi = {https://doi.org/10.1016/j.parco.2025.103125},
  author = {Casanova, Henri and Giersch, Arnaud and Legrand, Arnaud and Quinson, Martin and Suter, Fr{\'e}d{\'e}ric}
}

@article{logp,
author = {Culler, David and Karp, Richard and Patterson, David and Sahay, Abhijit and Schauser, Klaus Erik and Santos, Eunice and Subramonian, Ramesh and von Eicken, Thorsten},
title = {LogP: towards a realistic model of parallel computation},
year = {1993},
issue_date = {July 1993},
publisher = {Association for Computing Machinery},
address = {New York, NY, USA},
volume = {28},
number = {7},
issn = {0362-1340},
url = {https://doi.org/10.1145/173284.155333},
doi = {10.1145/173284.155333},
journal = {SIGPLAN Not.},
month = jul,
pages = {1–12},
numpages = {12}
}

@article{loggp,
title = {LogGP: Incorporating Long Messages into the LogP Model for Parallel Computation},
journal = {Journal of Parallel and Distributed Computing},
volume = {44},
number = {1},
pages = {71-79},
year = {1997},
issn = {0743-7315},
doi = {https://doi.org/10.1006/jpdc.1997.1346},
url = {https://www.sciencedirect.com/science/article/pii/S0743731597913460},
author = {Albert Alexandrov and Mihai F. Ionescu and Klaus E. Schauser and Chris Scheiman}
}

@inproceedings{plogp,
author = {Kielmann, Thilo and Bal, Henri E. and Verstoep, Kees},
title = {Fast Measurement of LogP Parameters for Message Passing Platforms},
year = {2000},
isbn = {354067442X},
publisher = {Springer-Verlag},
address = {Berlin, Heidelberg},
booktitle = {Proceedings of the 15 IPDPS 2000 Workshops on Parallel and Distributed Processing},
pages = {1176–1183},
numpages = {8},
series = {IPDPS '00}
}

@article{loggps,
author = {Ino, Fumihiko and Fujimoto, Noriyuki and Hagihara, Kenichi},
title = {LogGPS: a parallel computational model for synchronization analysis},
year = {2001},
issue_date = {July 2001},
publisher = {Association for Computing Machinery},
address = {New York, NY, USA},
volume = {36},
number = {7},
issn = {0362-1340},
url = {https://doi.org/10.1145/568014.379592},
doi = {10.1145/568014.379592},
journal = {SIGPLAN Not.},
month = jun,
pages = {133–142},
numpages = {10}
}

@article{hock_model,
title = {The communication challenge for MPP: Intel Paragon and Meiko CS-2},
journal = {Parallel Computing},
volume = {20},
number = {3},
pages = {389-398},
year = {1994},
issn = {0167-8191},
doi = {https://doi.org/10.1016/S0167-8191(06)80021-9},
url = {https://www.sciencedirect.com/science/article/pii/S0167819106800219},
author = {Roger W. Hockney}
}

@article{MPICH,
author = {Thakur, Rajeev and Rabenseifner, Rolf and Gropp, William},
year = {2005},
month = {01},
pages = {49-66},
title = {Optimization of Collective Communication Operations in MPICH.},
volume = {19},
journal = {IJHPCA}
}

@article{nuriyev2022model,
  title={Model-based selection of optimal MPI broadcast algorithms for multi-core clusters},
  author={Nuriyev, Emin and Rico-Gallego, Juan-Antonio and Lastovetsky, Alexey},
  journal={Journal of Parallel and Distributed Computing},
  volume={165},
  pages={1--16},
  year={2022},
  publisher={Elsevier}
}

@article{f7a8565e2a44463d956da9c22c088375,
title = "Improving the performance of collective operations in MPICH",
author = "Rajeev Thakur and Gropp, {William D.}",
year = "2003",
doi = "10.1007/978-3-540-39924-7_38",
language = "English (US)",
volume = "2840",
pages = "257--267",
journal = "Lecture Notes in Computer Science (including subseries Lecture Notes in Artificial Intelligence and Lecture Notes in Bioinformatics)",
issn = "0302-9743",
publisher = "Springer",
}

@article{flat_butterfly,
author = {Kim, John and Dally, William J. and Abts, Dennis},
title = {Flattened butterfly: a cost-efficient topology for high-radix networks},
year = {2007},
issue_date = {May 2007},
publisher = {Association for Computing Machinery},
address = {New York, NY, USA},
volume = {35},
number = {2},
issn = {0163-5964},
url = {https://doi.org/10.1145/1273440.1250679},
doi = {10.1145/1273440.1250679},
journal = {SIGARCH Comput. Archit. News},
month = jun,
pages = {126–137},
numpages = {12}
}

@INPROCEEDINGS{9926123,
  author={Jain, Nikhil and Bhatele, Abhinav and Howell, Louis H. and Böhme, David and Karlin, Ian and León, Edgar A. and Mubarak, Misbah and Wolfe, Noah and Gamblin, Todd and Leininger, Matthew L.},
  booktitle={SC17: International Conference for High Performance Computing, Networking, Storage and Analysis}, 
  title={Predicting the Performance Impact of Different Fat-Tree Configurations}, 
  year={2017},
  volume={},
  number={},
  pages={1-13},
  doi={}}

@article{almeida2025assessing,
  author = {Almeida, F. and Okon, E.},
  title = {Assessing the impact of high-performance computing on digital transformation: benefits, challenges, and size-dependent differences},
  journal = {The Journal of Supercomputing},
  volume = {81},
  pages = {795},
  year = {2025},
  doi = {10.1007/s11227-025-07281-z},
  url = {https://doi.org/10.1007/s11227-025-07281-z}
}

@article{HASANOV201530,
title = {Topology-oblivious optimization of MPI broadcast algorithms on extreme-scale platforms},
journal = {Simulation Modelling Practice and Theory},
volume = {58},
pages = {30-39},
year = {2015},
note = {Special Issue on TECHNIQUES AND APPLICATIONS FOR SUSTAINABLE ULTRASCALE COMPUTING SYSTEMS},
issn = {1569-190X},
doi = {https://doi.org/10.1016/j.simpat.2015.03.005},
url = {https://www.sciencedirect.com/science/article/pii/S1569190X15000465},
author = {Khalid Hasanov and Jean-No√´l Quintin and Alexey Lastovetsky}
}

@ARTICLE{8361902,
  author={Silvestre, Daniel and Hespanha, João P. and Silvestre, Carlos},
  journal={IEEE Transactions on Control of Network Systems}, 
  title={Broadcast and Gossip Stochastic Average Consensus Algorithms in Directed Topologies}, 
  year={2019},
  volume={6},
  number={2},
  pages={474-486},
  doi={10.1109/TCNS.2018.2839341}}

@INPROCEEDINGS{6337589,
  author={García, Marina and Vallejo, Enrique and Beivide, Ramón and Odriozola, Miguel and Camarero, Cristóbal and Valero, Mateo and Rodríguez, Germ'n and Labarta, Jesús and Minkenberg, Cyriel},
  booktitle={2012 41st International Conference on Parallel Processing}, 
  title={On-the-Fly Adaptive Routing in High-Radix Hierarchical Networks}, 
  year={2012},
  volume={},
  number={},
  pages={279-288},
  doi={10.1109/ICPP.2012.46}}

@article{article_pfDeng,
author = {Zhang, Peng and Deng, Yuefan},
year = {2012},
month = {12},
pages = {2245-2253},
title = {Design and Analysis of Pipelined Broadcast Algorithms for the All-Port Interlaced Bypass Torus Networks},
volume = {23},
journal = {Parallel and Distributed Systems, IEEE Transactions on},
doi = {10.1109/TPDS.2012.93}
}

@article{dongarra2003linpack,
  author = {Dongarra, J. J. and Luszczek, P. and Petitet, A.},
  title = {The {LINPACK} Benchmark: Past, Present and Future},
  journal = {Concurrency and Computation: Practice and Experience},
  volume = {15},
  pages = {803--820},
  year = {2003},
  doi = {10.1002/cpe.728}
}

@inproceedings{10.1145/1400751.1400773,
author = {Berenbrink, Petra and Elsaesser, Robert and Friedetzky, Tom},
title = {Efficient randomised broadcasting in random regular networks with applications in peer-to-peer systems},
year = {2008},
isbn = {9781595939890},
publisher = {Association for Computing Machinery},
address = {New York, NY, USA},
url = {https://doi.org/10.1145/1400751.1400773},
doi = {10.1145/1400751.1400773},
booktitle = {Proceedings of the Twenty-Seventh ACM Symposium on Principles of Distributed Computing},
pages = {155–164},
numpages = {10},
location = {Toronto, Canada},
series = {PODC '08}
}

@inproceedings{10.1145/800133.804346,
author = {Gabow, Harold N. and Kariv, Oded},
title = {Algorithms for edge coloring bipartite graphs},
year = {1978},
isbn = {9781450374378},
publisher = {Association for Computing Machinery},
address = {New York, NY, USA},
url = {https://doi.org/10.1145/800133.804346},
doi = {10.1145/800133.804346},
booktitle = {Proceedings of the Tenth Annual ACM Symposium on Theory of Computing},
pages = {184–192},
numpages = {9},
location = {San Diego, California, USA},
series = {STOC '78}
}

@article{article1212,
author = {Sinha, Koushik and Srimani, Pradip},
year = {2006},
month = {08},
pages = {115-144},
title = {Deterministic Broadcast and Gossiping Algorithms for Ad hoc Networks},
volume = {37},
journal = {The Journal of Supercomputing},
doi = {10.1007/s11227-006-6255-3}
}

@InProceedings{10.1007/978-3-540-30218-6_28,
author="Tr{\"a}ff, Jesper Larsson",
editor="Kranzlm{\"u}ller, Dieter
and Kacsuk, P{\'e}ter
and Dongarra, Jack",
title="A Simple Work-Optimal Broadcast Algorithm for Message-Passing Parallel Systems",
booktitle="Recent Advances in Parallel Virtual Machine and Message Passing Interface",
year="2004",
publisher="Springer Berlin Heidelberg",
address="Berlin, Heidelberg",
pages="173--180",
isbn="978-3-540-30218-6"
}

@ARTICLE{679219,
  author={Louri, A. and Weech, B. and Neocleous, C.},
  journal={IEEE Transactions on Parallel and Distributed Systems}, 
  title={A spanning multichannel linked hypercube: a gradually scalable optical interconnection network for massively parallel computing}, 
  year={1998},
  volume={9},
  number={5},
  pages={497-512},
  doi={10.1109/71.679219}}

@InProceedings{10.1007/11557654_8,
author="Tr{\"a}ff, Jesper Larsson
and Ripke, Andreas",
editor="Yang, Laurence T.
and Rana, Omer F.
and Di Martino, Beniamino
and Dongarra, Jack",
title="Optimal Broadcast for Fully Connected Networks",
booktitle="High Performance Computing and Communications",
year="2005",
publisher="Springer Berlin Heidelberg",
address="Berlin, Heidelberg",
pages="45--56",
isbn="978-3-540-32079-1"
}

@INPROCEEDINGS{9355242,
  author={Jia, Weile and Wang, Han and Chen, Mohan and Lu, Denghui and Lin, Lin and Car, Roberto and Weinan, E and Zhang, Linfeng},
  booktitle={SC20: International Conference for High Performance Computing, Networking, Storage and Analysis}, 
  title={Pushing the Limit of Molecular Dynamics with Ab Initio Accuracy to 100 Million Atoms with Machine Learning}, 
  year={2020},
  volume={},
  number={},
  pages={1-14},
  doi={10.1109/SC41405.2020.00009}}

@article{doi:10.1177/10943420231183688,
author = {Jerry Watkins and Max Carlson and Kyle Shan and Irina Tezaur and Mauro Perego and Luca Bertagna and Carolyn Kao and Matthew J Hoffman and Stephen F Price},
title ={Performance portable ice-sheet modeling with MALI},

journal = {The International Journal of High Performance Computing Applications},
volume = {37},
number = {5},
pages = {600-625},
year = {2023},
doi = {10.1177/10943420231183688},

URL = { 
    
        https://doi.org/10.1177/10943420231183688
    
    

}
}

@article{doi:10.1021/ct9000685,
author = {Harvey, M. J. and Giupponi, G. and Fabritiis, G. De},
title = {ACEMD: Accelerating Biomolecular Dynamics in the Microsecond Time Scale},
journal = {Journal of Chemical Theory and Computation},
volume = {5},
number = {6},
pages = {1632-1639},
year = {2009},
doi = {10.1021/ct9000685},
    note ={PMID: 26609855},

URL = { 
    
        https://doi.org/10.1021/ct9000685
    
    

},
eprint = { 
    
        https://doi.org/10.1021/ct9000685
    
    

}

}

@Article{app10196717,
AUTHOR = {Woo, Junghoon and Choi, Hyeonseong and Lee, Jaehwan},
TITLE = {Empirical Performance Analysis of Collective Communication for Distributed Deep Learning in a Many-Core CPU Environment},
JOURNAL = {Applied Sciences},
VOLUME = {10},
YEAR = {2020},
NUMBER = {19},
ARTICLE-NUMBER = {6717},
URL = {https://www.mdpi.com/2076-3417/10/19/6717},
ISSN = {2076-3417},
DOI = {10.3390/app10196717}
}

@InProceedings{10.1007/11549468_87,
author="Eleftheriou, Maria
and Fitch, Blake
and Rayshubskiy, Aleksandr
and Ward, T. J. Christopher
and Germain, Robert",
editor="Cunha, Jos{\'e} C.
and Medeiros, Pedro D.",
title="Performance Measurements of the 3D FFT on the Blue Gene/L Supercomputer",
booktitle="Euro-Par 2005 Parallel Processing",
year="2005",
publisher="Springer Berlin Heidelberg",
address="Berlin, Heidelberg",
pages="795--803",
isbn="978-3-540-31925-2"
}

@techreport{10.5555/899255,
author = {Mitra, Prasenjit and Payne, David and Shuler, Lance and van de Geijn, Robert and Watts, Jerrell},
title = {Fast Collective Communication Libraries, Please},
year = {1995},
publisher = {University of Texas at Austin},
address = {USA}
}

@INPROCEEDINGS{1420226,
  author={Pjesivac-Grbovic, J. and Angskun, T. and Bosilca, G. and Fagg, G.E. and Gabriel, E. and Dongarra, J.J.},
  booktitle={19th IEEE International Parallel and Distributed Processing Symposium}, 
  title={Performance analysis of MPI collective operations}, 
  year={2005},
  volume={},
  number={},
  pages={8 pp.-},
  doi={10.1109/IPDPS.2005.335}}

@article{BDST,
author = {Angel, Omer and Flaxman, Abraham D. and Wilson, David B.},
title = {A sharp threshold for minimum bounded-depth and bounded-diameter spanning trees and Steiner trees in random networks},
year = {2012},
issue_date = {January 2012},
publisher = {Springer-Verlag},
address = {Berlin, Heidelberg},
volume = {32},
number = {1},
issn = {0209-9683},
url = {https://doi.org/10.1007/s00493-012-2552-z},
doi = {10.1007/s00493-012-2552-z},
journal = {Combinatorica},
month = jan,
pages = {1–33},
numpages = {33}
}

\newpage
\begin{appendices}

\section{Timing Figures}\label{time_figs}

\begin{figure}[H]
\centering
\includegraphics[width=\linewidth]{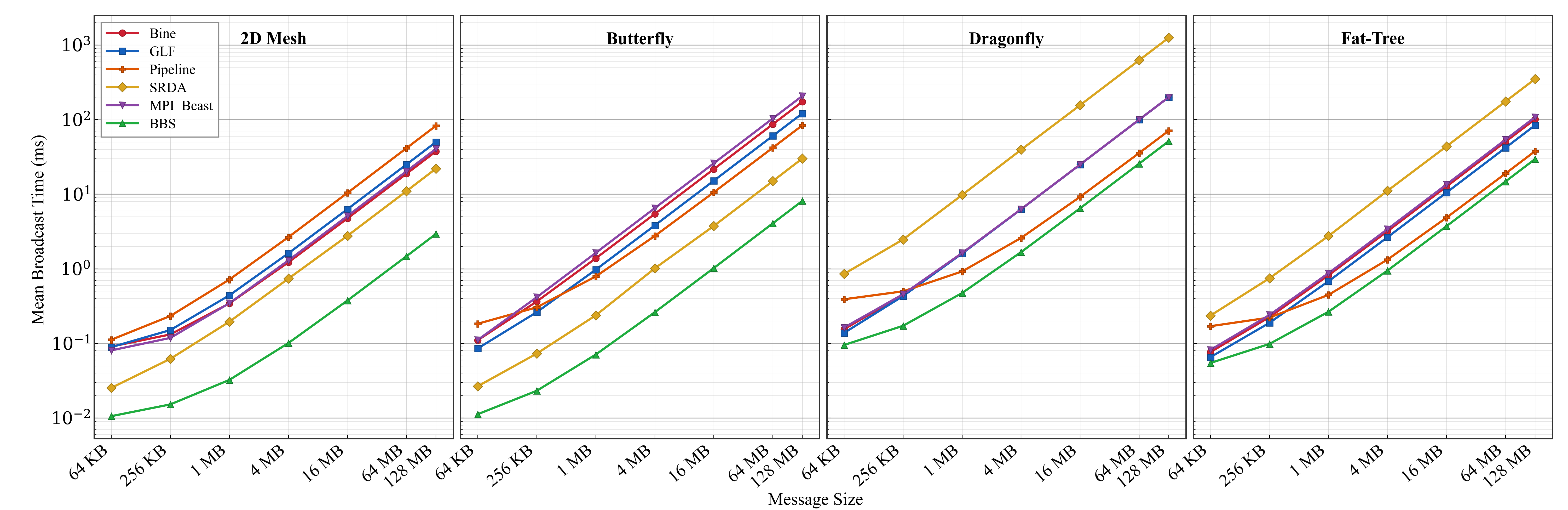}
\caption{$T_{\text{avg}}$ across various message sizes on 128-node topologies.}
\label{time_128}
\end{figure}

\begin{figure}[H]
\centering
\includegraphics[width=\linewidth]{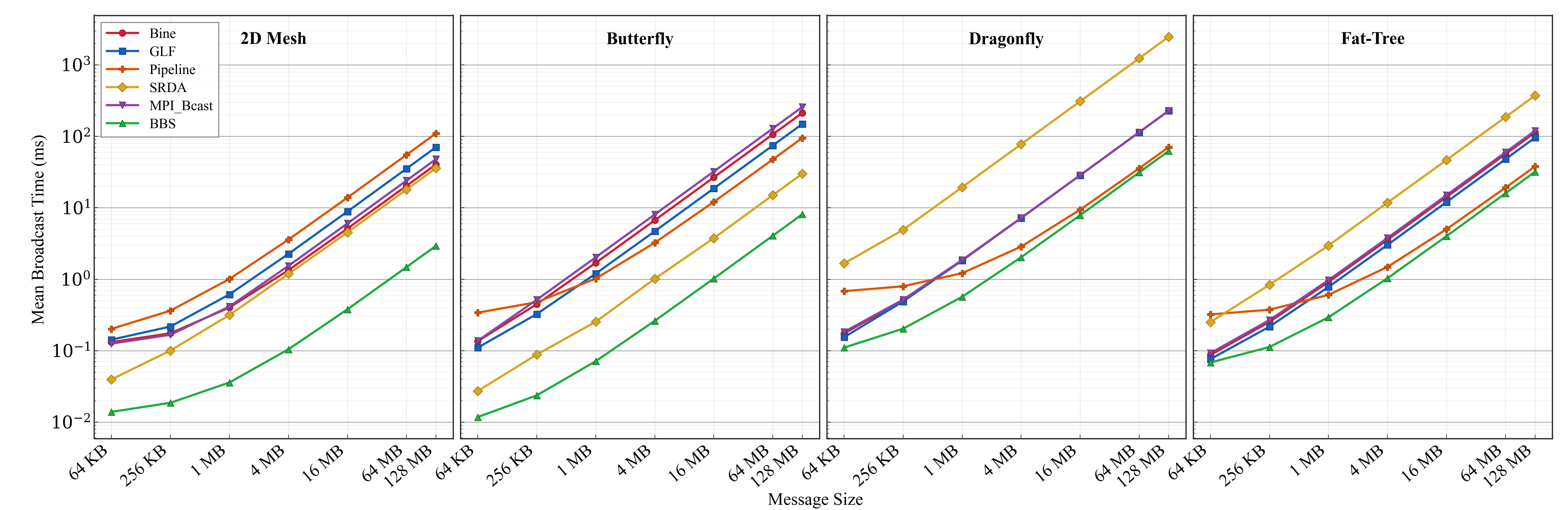}
\caption{$T_{\text{avg}}$ across various message sizes on 256-node topologies.}
\label{time_256}
\end{figure}

\begin{figure}[H]
\centering
\includegraphics[width=\linewidth]{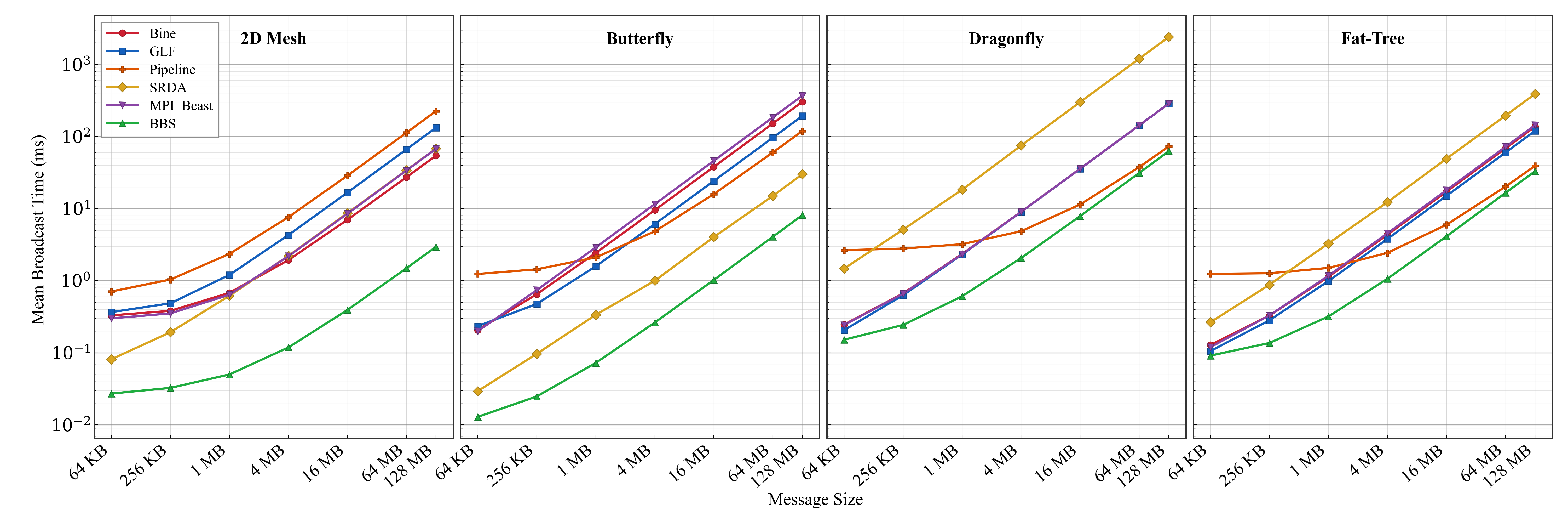}
\caption{$T_{\text{avg}}$ across various message sizes on 1024-node topologies.}
\label{time_1024}
\end{figure}

\section{Performance Tables}\label{time_tables}

\begin{tableorg}[htbp]
\centering
\caption{Broadcast statistics for 2D the Mesh and Butterfly topologies with 128 nodes. The mean, maximum, minimum, and standard deviation are reported.}
\label{tab:results_N128_nonrouter}
\setlength{\tabcolsep}{6pt}
\renewcommand{\arraystretch}{1.00}
\resizebox{!}{0.48\textheight}{%
\begin{tabular}{|l|c|c|c|c|c|c|c|}
\hline
\multicolumn{8}{|c|}{\textbf{2D Mesh}} \\
\hline
Message Size & Metric & BBS & Bine & GLF & Pipeline & SRDA & MPI\_Bcast \\
\hline
64 KB &
\makecell[c]{$T_{\text{avg}}$\\$T_{\max}$\\$T_{\min}$\\$\sigma$} &
\makecell[c]{\textbf{10.6 \textmu s}\\16.0 \textmu s\\6.98 \textmu s\\2.11 \textmu s} &
\makecell[c]{90.5 \textmu s\\115 \textmu s\\47.6 \textmu s\\17.4 \textmu s} &
\makecell[c]{88.9 \textmu s\\134 \textmu s\\53.7 \textmu s\\19.1 \textmu s} &
\makecell[c]{112 \textmu s\\114 \textmu s\\94.8 \textmu s\\2.23 \textmu s} &
\makecell[c]{25.3 \textmu s\\26.8 \textmu s\\23.9 \textmu s\\681 ns} &
\makecell[c]{80.4 \textmu s\\124 \textmu s\\36.7 \textmu s\\20.1 \textmu s} \\
\hline
256 KB &
\makecell[c]{$T_{\text{avg}}$\\$T_{\max}$\\$T_{\min}$\\$\sigma$} &
\makecell[c]{\textbf{15.2 \textmu s}\\21.5 \textmu s\\11.3 \textmu s\\2.20 \textmu s} &
\makecell[c]{132 \textmu s\\159 \textmu s\\78.3 \textmu s\\21.5 \textmu s} &
\makecell[c]{151 \textmu s\\203 \textmu s\\92.5 \textmu s\\29.0 \textmu s} &
\makecell[c]{233 \textmu s\\235 \textmu s\\183 \textmu s\\4.68 \textmu s} &
\makecell[c]{62.0 \textmu s\\66.2 \textmu s\\58.8 \textmu s\\1.71 \textmu s} &
\makecell[c]{118 \textmu s\\168 \textmu s\\67.4 \textmu s\\22.5 \textmu s} \\
\hline
1 MB &
\makecell[c]{$T_{\text{avg}}$\\$T_{\max}$\\$T_{\min}$\\$\sigma$} &
\makecell[c]{\textbf{32.3 \textmu s}\\38.6 \textmu s\\28.5 \textmu s\\2.21 \textmu s} &
\makecell[c]{343 \textmu s\\419 \textmu s\\201 \textmu s\\61.2 \textmu s} &
\makecell[c]{440 \textmu s\\624 \textmu s\\266 \textmu s\\86.6 \textmu s} &
\makecell[c]{717 \textmu s\\722 \textmu s\\536 \textmu s\\16.4 \textmu s} &
\makecell[c]{195 \textmu s\\205 \textmu s\\188 \textmu s\\3.64 \textmu s} &
\makecell[c]{345 \textmu s\\496 \textmu s\\190 \textmu s\\68.7 \textmu s} \\
\hline
\end{tabular}%
}
\end{tableorg}

\clearpage
\begin{tableorg}[htbp]
\centering
\caption{Broadcast statistics for Dragonfly and Fat-Tree topologies with 128 nodes. The mean, maximum, minimum, and standard deviation are reported.}
\label{tab:results_N128_router}
\setlength{\tabcolsep}{6pt}
\renewcommand{\arraystretch}{1.00}
\resizebox{!}{0.48\textheight}{%
\begin{tabular}{|l|c|c|c|c|c|c|c|}
\hline
\multicolumn{8}{|c|}{\textbf{Dragonfly}} \\
\hline
Message Size & Metric & BBS & Bine & GLF & Pipeline & SRDA & MPI\_Bcast \\
\hline
64 KB &
\makecell[c]{$T_{\text{avg}}$\\$T_{\max}$\\$T_{\min}$\\$\sigma$} &
\makecell[c]{\textbf{94.9 \textmu s}\\95.5 \textmu s\\93.6 \textmu s\\564 ns} &
\makecell[c]{154 \textmu s\\162 \textmu s\\145 \textmu s\\4.80 \textmu s} &
\makecell[c]{138 \textmu s\\141 \textmu s\\131 \textmu s\\2.72 \textmu s} &
\makecell[c]{390 \textmu s\\400 \textmu s\\377 \textmu s\\4.77 \textmu s} &
\makecell[c]{853 \textmu s\\854 \textmu s\\851 \textmu s\\717 ns} &
\makecell[c]{163 \textmu s\\186 \textmu s\\131 \textmu s\\12.6 \textmu s} \\
\hline
256 KB &
\makecell[c]{$T_{\text{avg}}$\\$T_{\max}$\\$T_{\min}$\\$\sigma$} &
\makecell[c]{\textbf{172 \textmu s}\\172 \textmu s\\171 \textmu s\\501 ns} &
\makecell[c]{447 \textmu s\\455 \textmu s\\438 \textmu s\\4.80 \textmu s} &
\makecell[c]{430 \textmu s\\433 \textmu s\\424 \textmu s\\2.72 \textmu s} &
\makecell[c]{498 \textmu s\\517 \textmu s\\485 \textmu s\\5.53 \textmu s} &
\makecell[c]{2.47 ms\\2.47 ms\\2.47 ms\\1.05 \textmu s} &
\makecell[c]{455 \textmu s\\478 \textmu s\\424 \textmu s\\12.6 \textmu s} \\
\hline
1 MB &
\makecell[c]{$T_{\text{avg}}$\\$T_{\max}$\\$T_{\min}$\\$\sigma$} &
\makecell[c]{\textbf{473 \textmu s}\\473 \textmu s\\471 \textmu s\\654 ns} &
\makecell[c]{1.62 ms\\1.63 ms\\1.61 ms\\4.80 \textmu s} &
\makecell[c]{1.60 ms\\1.60 ms\\1.59 ms\\2.72 \textmu s} &
\makecell[c]{921 \textmu s\\990 \textmu s\\894 \textmu s\\21.0 \textmu s} &
\makecell[c]{9.76 ms\\9.77 ms\\9.76 ms\\1.96 \textmu s} &
\makecell[c]{1.63 ms\\1.65 ms\\1.59 ms\\12.6 \textmu s} \\
\hline
\end{tabular}%
}
\end{tableorg}

\clearpage
\begin{tableorg}[htbp]
\centering
\caption{Broadcast statistics for 2D Mesh and Butterfly topologies with 256 nodes. The mean, maximum, minimum, and standard deviation are reported.}
\label{tab:results_N256_nonrouter}
\setlength{\tabcolsep}{6pt}
\renewcommand{\arraystretch}{1.00}
\resizebox{!}{0.48\textheight}{%
\begin{tabular}{|l|c|c|c|c|c|c|c|}
\hline
\multicolumn{8}{|c|}{\textbf{2D Mesh}} \\
\hline
Message Size & Metric & BBS & Bine & GLF & Pipeline & SRDA & MPI\_Bcast \\
\hline
64 KB &
\makecell[c]{$T_{\text{avg}}$\\$T_{\max}$\\$T_{\min}$\\$\sigma$} &
\makecell[c]{\textbf{13.9 \textmu s}\\24.0 \textmu s\\7.39 \textmu s\\3.37 \textmu s} &
\makecell[c]{132 \textmu s\\169 \textmu s\\62.3 \textmu s\\24.5 \textmu s} &
\makecell[c]{142 \textmu s\\213 \textmu s\\96.9 \textmu s\\23.8 \textmu s} &
\makecell[c]{201 \textmu s\\205 \textmu s\\171 \textmu s\\3.08 \textmu s} &
\makecell[c]{39.5 \textmu s\\41.8 \textmu s\\37.9 \textmu s\\976 ns} &
\makecell[c]{125 \textmu s\\203 \textmu s\\48.3 \textmu s\\31.4 \textmu s} \\
\hline
256 KB &
\makecell[c]{$T_{\text{avg}}$\\$T_{\max}$\\$T_{\min}$\\$\sigma$} &
\makecell[c]{\textbf{18.6 \textmu s}\\28.7 \textmu s\\12.1 \textmu s\\3.45 \textmu s} &
\makecell[c]{175 \textmu s\\217 \textmu s\\97.4 \textmu s\\27.8 \textmu s} &
\makecell[c]{217 \textmu s\\314 \textmu s\\160 \textmu s\\27.3 \textmu s} &
\makecell[c]{362 \textmu s\\366 \textmu s\\259 \textmu s\\6.86 \textmu s} &
\makecell[c]{100 \textmu s\\107 \textmu s\\95.1 \textmu s\\2.97 \textmu s} &
\makecell[c]{166 \textmu s\\250 \textmu s\\83.4 \textmu s\\33.3 \textmu s} \\
\hline
1 MB &
\makecell[c]{$T_{\text{avg}}$\\$T_{\max}$\\$T_{\min}$\\$\sigma$} &
\makecell[c]{\textbf{35.8 \textmu s}\\45.8 \textmu s\\29.3 \textmu s\\3.50 \textmu s} &
\makecell[c]{401 \textmu s\\478 \textmu s\\238 \textmu s\\60.8 \textmu s} &
\makecell[c]{613 \textmu s\\868 \textmu s\\481 \textmu s\\65.0 \textmu s} &
\makecell[c]{1.01 ms\\1.01 ms\\612 \textmu s\\24.8 \textmu s} &
\makecell[c]{316 \textmu s\\326 \textmu s\\304 \textmu s\\4.94 \textmu s} &
\makecell[c]{415 \textmu s\\610 \textmu s\\224 \textmu s\\78.6 \textmu s} \\
\hline
\end{tabular}%
}
\end{tableorg}

\clearpage
\begin{tableorg}[htbp]
\centering
\caption{Broadcast statistics for Dragonfly and Fat-Tree topologies with 256 nodes. The mean, maximum, minimum, and standard deviation are reported.}
\label{tab:results_N256_router}
\setlength{\tabcolsep}{6pt}
\renewcommand{\arraystretch}{1.00}
\resizebox{!}{0.48\textheight}{%
\begin{tabular}{|l|c|c|c|c|c|c|c|}
\hline
\multicolumn{8}{|c|}{\textbf{Dragonfly}} \\
\hline
Message Size & Metric & BBS & Bine & GLF & Pipeline & SRDA & MPI\_Bcast \\
\hline
64 KB &
\makecell[c]{$T_{\text{avg}}$\\$T_{\max}$\\$T_{\min}$\\$\sigma$} &
\makecell[c]{\textbf{110 \textmu s}\\111 \textmu s\\102 \textmu s\\1.07 \textmu s} &
\makecell[c]{174 \textmu s\\187 \textmu s\\162 \textmu s\\5.67 \textmu s} &
\makecell[c]{154 \textmu s\\157 \textmu s\\148 \textmu s\\2.76 \textmu s} &
\makecell[c]{680 \textmu s\\708 \textmu s\\635 \textmu s\\14.5 \textmu s} &
\makecell[c]{1.66 ms\\1.66 ms\\1.66 ms\\688 ns} &
\makecell[c]{183 \textmu s\\210 \textmu s\\148 \textmu s\\13.3 \textmu s} \\
\hline
256 KB &
\makecell[c]{$T_{\text{avg}}$\\$T_{\max}$\\$T_{\min}$\\$\sigma$} &
\makecell[c]{\textbf{203 \textmu s}\\204 \textmu s\\185 \textmu s\\2.08 \textmu s} &
\makecell[c]{508 \textmu s\\522 \textmu s\\496 \textmu s\\5.67 \textmu s} &
\makecell[c]{489 \textmu s\\491 \textmu s\\482 \textmu s\\2.76 \textmu s} &
\makecell[c]{796 \textmu s\\815 \textmu s\\757 \textmu s\\10.1 \textmu s} &
\makecell[c]{4.92 ms\\4.92 ms\\4.91 ms\\1.45 \textmu s} &
\makecell[c]{517 \textmu s\\545 \textmu s\\482 \textmu s\\13.3 \textmu s} \\
\hline
1 MB &
\makecell[c]{$T_{\text{avg}}$\\$T_{\max}$\\$T_{\min}$\\$\sigma$} &
\makecell[c]{\textbf{568 \textmu s}\\569 \textmu s\\538 \textmu s\\3.31 \textmu s} &
\makecell[c]{1.85 ms\\1.86 ms\\1.83 ms\\5.67 \textmu s} &
\makecell[c]{1.83 ms\\1.83 ms\\1.82 ms\\2.76 \textmu s} &
\makecell[c]{1.21 ms\\1.27 ms\\1.18 ms\\16.2 \textmu s} &
\makecell[c]{19.4 ms\\19.4 ms\\19.4 ms\\1.89 \textmu s} &
\makecell[c]{1.86 ms\\1.88 ms\\1.82 ms\\13.3 \textmu s} \\
\hline
\end{tabular}%
}
\end{tableorg}

\clearpage
\begin{tableorg}[htbp]
\centering
\caption{Broadcast statistics for 2D Mesh and Butterfly topologies with 512 nodes. The mean, maximum, minimum, and standard deviation are reported.}
\label{tab:results_N512_nonrouter}
\setlength{\tabcolsep}{6pt}
\renewcommand{\arraystretch}{1.00}
\resizebox{!}{0.48\textheight}{%
\begin{tabular}{|l|c|c|c|c|c|c|c|}
\hline
\multicolumn{8}{|c|}{\textbf{2D Mesh}} \\
\hline
Message Size & Metric & BBS & Bine & GLF & Pipeline & SRDA & MPI\_Bcast \\
\hline
64 KB &
\makecell[c]{$T_{\text{avg}}$\\$T_{\max}$\\$T_{\min}$\\$\sigma$} &
\makecell[c]{\textbf{20.9 \textmu s}\\35.1 \textmu s\\12.9 \textmu s\\4.41 \textmu s} &
\makecell[c]{221 \textmu s\\609 \textmu s\\109 \textmu s\\45.3 \textmu s} &
\makecell[c]{227 \textmu s\\566 \textmu s\\127 \textmu s\\52.5 \textmu s} &
\makecell[c]{378 \textmu s\\380 \textmu s\\338 \textmu s\\2.91 \textmu s} &
\makecell[c]{50.9 \textmu s\\683 \textmu s\\46.9 \textmu s\\28.0 \textmu s} &
\makecell[c]{192 \textmu s\\316 \textmu s\\68.2 \textmu s\\49.6 \textmu s} \\
\hline
256 KB &
\makecell[c]{$T_{\text{avg}}$\\$T_{\max}$\\$T_{\min}$\\$\sigma$} &
\makecell[c]{\textbf{26.2 \textmu s}\\40.1 \textmu s\\18.0 \textmu s\\4.51 \textmu s} &
\makecell[c]{273 \textmu s\\638 \textmu s\\149 \textmu s\\48.9 \textmu s} &
\makecell[c]{330 \textmu s\\630 \textmu s\\195 \textmu s\\65.4 \textmu s} &
\makecell[c]{626 \textmu s\\628 \textmu s\\510 \textmu s\\5.65 \textmu s} &
\makecell[c]{113 \textmu s\\723 \textmu s\\106 \textmu s\\27.2 \textmu s} &
\makecell[c]{237 \textmu s\\369 \textmu s\\108 \textmu s\\50.5 \textmu s} \\
\hline
1 MB &
\makecell[c]{$T_{\text{avg}}$\\$T_{\max}$\\$T_{\min}$\\$\sigma$} &
\makecell[c]{\textbf{43.4 \textmu s}\\57.5 \textmu s\\35.1 \textmu s\\4.51 \textmu s} &
\makecell[c]{544 \textmu s\\985 \textmu s\\314 \textmu s\\90.1 \textmu s} &
\makecell[c]{878 \textmu s\\1.34 ms\\544 \textmu s\\169 \textmu s} &
\makecell[c]{1.62 ms\\1.62 ms\\1.20 ms\\19.0 \textmu s} &
\makecell[c]{351 \textmu s\\958 \textmu s\\337 \textmu s\\27.6 \textmu s} &
\makecell[c]{523 \textmu s\\773 \textmu s\\266 \textmu s\\97.7 \textmu s} \\
\hline
\end{tabular}%
}
\end{tableorg}

\clearpage
\begin{tableorg}[htbp]
\centering
\caption{Broadcast statistics for Dragonfly and Fat-Tree topologies with 512 nodes. The mean, maximum, minimum, and standard deviation are reported.}
\label{tab:results_N512_router}
\setlength{\tabcolsep}{6pt}
\renewcommand{\arraystretch}{1.00}
\resizebox{!}{0.48\textheight}{%
\begin{tabular}{|l|c|c|c|c|c|c|c|}
\hline
\multicolumn{8}{|c|}{\textbf{Dragonfly}} \\
\hline
Message Size & Metric & BBS & Bine & GLF & Pipeline & SRDA & MPI\_Bcast \\
\hline
64 KB &
\makecell[c]{$T_{\text{avg}}$\\$T_{\max}$\\$T_{\min}$\\$\sigma$} &
\makecell[c]{\textbf{131 \textmu s}\\131 \textmu s\\121 \textmu s\\1.14 \textmu s} &
\makecell[c]{202 \textmu s\\569 \textmu s\\186 \textmu s\\17.5 \textmu s} &
\makecell[c]{180 \textmu s\\533 \textmu s\\168 \textmu s\\16.1 \textmu s} &
\makecell[c]{1.33 ms\\1.38 ms\\1.18 ms\\30.2 \textmu s} &
\makecell[c]{1.59 ms\\2.21 ms\\1.58 ms\\27.8 \textmu s} &
\makecell[c]{210 \textmu s\\575 \textmu s\\168 \textmu s\\22.2 \textmu s} \\
\hline
256 KB &
\makecell[c]{$T_{\text{avg}}$\\$T_{\max}$\\$T_{\min}$\\$\sigma$} &
\makecell[c]{\textbf{223 \textmu s}\\224 \textmu s\\205 \textmu s\\1.78 \textmu s} &
\makecell[c]{579 \textmu s\\948 \textmu s\\563 \textmu s\\17.5 \textmu s} &
\makecell[c]{560 \textmu s\\917 \textmu s\\548 \textmu s\\16.3 \textmu s} &
\makecell[c]{1.46 ms\\1.48 ms\\1.41 ms\\12.5 \textmu s} &
\makecell[c]{5.24 ms\\5.71 ms\\5.21 ms\\21.8 \textmu s} &
\makecell[c]{586 \textmu s\\615 \textmu s\\545 \textmu s\\15.2 \textmu s} \\
\hline
1 MB &
\makecell[c]{$T_{\text{avg}}$\\$T_{\max}$\\$T_{\min}$\\$\sigma$} &
\makecell[c]{\textbf{588 \textmu s}\\589 \textmu s\\543 \textmu s\\3.55 \textmu s} &
\makecell[c]{2.09 ms\\2.46 ms\\2.07 ms\\17.5 \textmu s} &
\makecell[c]{2.07 ms\\2.43 ms\\2.06 ms\\16.5 \textmu s} &
\makecell[c]{1.88 ms\\1.94 ms\\1.85 ms\\16.5 \textmu s} &
\makecell[c]{19.5 ms\\20.0 ms\\19.5 ms\\83.4 \textmu s} &
\makecell[c]{2.09 ms\\2.12 ms\\2.05 ms\\15.3 \textmu s} \\
\hline
\end{tabular}%
}
\end{tableorg}

\clearpage
\begin{tableorg}[htbp]
\centering
\caption{Broadcast statistics for 2D Mesh and Butterfly topologies with 1024 nodes. The mean, maximum, minimum, and standard deviation are reported.}
\label{tab:results_N1024_nonrouter}
\setlength{\tabcolsep}{6pt}
\renewcommand{\arraystretch}{1.00}
\resizebox{!}{0.48\textheight}{%
\begin{tabular}{|l|c|c|c|c|c|c|c|}
\hline
\multicolumn{8}{|c|}{\textbf{2D Mesh}} \\
\hline
Message Size & Metric & BBS & Bine & GLF & Pipeline & SRDA & MPI\_Bcast \\
\hline
64 KB &
\makecell[c]{$T_{\text{avg}}$\\$T_{\max}$\\$T_{\min}$\\$\sigma$} &
\makecell[c]{\textbf{27.2 \textmu s}\\53.2 \textmu s\\13.9 \textmu s\\7.01 \textmu s} &
\makecell[c]{330 \textmu s\\1.22 ms\\151 \textmu s\\66.0 \textmu s} &
\makecell[c]{366 \textmu s\\769 \textmu s\\235 \textmu s\\57.9 \textmu s} &
\makecell[c]{708 \textmu s\\711 \textmu s\\639 \textmu s\\3.47 \textmu s} &
\makecell[c]{81.3 \textmu s\\713 \textmu s\\76.8 \textmu s\\20.1 \textmu s} &
\makecell[c]{300 \textmu s\\508 \textmu s\\90.7 \textmu s\\74.1 \textmu s} \\
\hline
256 KB &
\makecell[c]{$T_{\text{avg}}$\\$T_{\max}$\\$T_{\min}$\\$\sigma$} &
\makecell[c]{\textbf{32.6 \textmu s}\\60.3 \textmu s\\18.7 \textmu s\\7.13 \textmu s} &
\makecell[c]{382 \textmu s\\933 \textmu s\\196 \textmu s\\63.6 \textmu s} &
\makecell[c]{485 \textmu s\\1.14 ms\\371 \textmu s\\67.9 \textmu s} &
\makecell[c]{1.04 ms\\1.04 ms\\811 \textmu s\\7.58 \textmu s} &
\makecell[c]{193 \textmu s\\811 \textmu s\\186 \textmu s\\19.7 \textmu s} &
\makecell[c]{352 \textmu s\\574 \textmu s\\136 \textmu s\\76.3 \textmu s} \\
\hline
1 MB &
\makecell[c]{$T_{\text{avg}}$\\$T_{\max}$\\$T_{\min}$\\$\sigma$} &
\makecell[c]{\textbf{50.0 \textmu s}\\77.3 \textmu s\\36.1 \textmu s\\7.23 \textmu s} &
\makecell[c]{679 \textmu s\\9.80 ms\\375 \textmu s\\300 \textmu s} &
\makecell[c]{1.20 ms\\1.80 ms\\906 \textmu s\\125 \textmu s} &
\makecell[c]{2.35 ms\\2.36 ms\\1.50 ms\\26.9 \textmu s} &
\makecell[c]{617 \textmu s\\1.20 ms\\594 \textmu s\\21.6 \textmu s} &
\makecell[c]{639 \textmu s\\976 \textmu s\\323 \textmu s\\117 \textmu s} \\
\hline
\end{tabular}%
}
\end{tableorg}

\clearpage
\begin{tableorg}[htbp]
\centering
\caption{Broadcast statistics for Dragonfly and Fat-Tree topologies with 1024 nodes. The mean, maximum, minimum, and standard deviation are reported.}
\label{tab:results_N1024_router}
\setlength{\tabcolsep}{6pt}
\renewcommand{\arraystretch}{1.00}
\resizebox{!}{0.48\textheight}{%
\begin{tabular}{|l|c|c|c|c|c|c|c|}
\hline
\multicolumn{8}{|c|}{\textbf{Dragonfly}} \\
\hline
Message Size & Metric & BBS & Bine & GLF & Pipeline & SRDA & MPI\_Bcast \\
\hline
64 KB &
\makecell[c]{$T_{\text{avg}}$\\$T_{\max}$\\$T_{\min}$\\$\sigma$} &
\makecell[c]{\textbf{152 \textmu s}\\152 \textmu s\\138 \textmu s\\1.48 \textmu s} &
\makecell[c]{246 \textmu s\\800 \textmu s\\225 \textmu s\\19.0 \textmu s} &
\makecell[c]{206 \textmu s\\569 \textmu s\\190 \textmu s\\12.6 \textmu s} &
\makecell[c]{2.65 ms\\2.72 ms\\2.23 ms\\60.3 \textmu s} &
\makecell[c]{1.47 ms\\1.99 ms\\1.46 ms\\16.7 \textmu s} &
\makecell[c]{243 \textmu s\\2.04 ms\\191 \textmu s\\59.0 \textmu s} \\
\hline
256 KB &
\makecell[c]{$T_{\text{avg}}$\\$T_{\max}$\\$T_{\min}$\\$\sigma$} &
\makecell[c]{\textbf{243 \textmu s}\\244 \textmu s\\224 \textmu s\\1.68 \textmu s} &
\makecell[c]{665 \textmu s\\1.23 ms\\647 \textmu s\\19.2 \textmu s} &
\makecell[c]{629 \textmu s\\1000 \textmu s\\615 \textmu s\\12.7 \textmu s} &
\makecell[c]{2.79 ms\\2.82 ms\\2.74 ms\\14.7 \textmu s} &
\makecell[c]{5.11 ms\\5.58 ms\\5.09 ms\\16.8 \textmu s} &
\makecell[c]{659 \textmu s\\691 \textmu s\\609 \textmu s\\17.6 \textmu s} \\
\hline
1 MB &
\makecell[c]{$T_{\text{avg}}$\\$T_{\max}$\\$T_{\min}$\\$\sigma$} &
\makecell[c]{\textbf{609 \textmu s}\\609 \textmu s\\560 \textmu s\\2.92 \textmu s} &
\makecell[c]{2.35 ms\\10.7 ms\\2.32 ms\\261 \textmu s} &
\makecell[c]{2.31 ms\\2.67 ms\\2.29 ms\\12.6 \textmu s} &
\makecell[c]{3.21 ms\\3.27 ms\\3.17 ms\\17.4 \textmu s} &
\makecell[c]{18.3 ms\\18.5 ms\\18.2 ms\\35.2 \textmu s} &
\makecell[c]{2.33 ms\\2.37 ms\\2.28 ms\\17.4 \textmu s} \\
\hline
\end{tabular}%
}
\end{tableorg}

\end{appendices}

\end{document}